\documentclass[preprint,5p,twocolumn,times]{elsarticle}

\usepackage{amsmath}
\usepackage{amssymb}
\usepackage{amsfonts}
\usepackage{amsthm}
\usepackage{mathtools}
\usepackage{mathrsfs}
\usepackage{xcolor}
\usepackage[hidelinks]{hyperref}

\theoremstyle{plain}

\theoremstyle{definition}

\theoremstyle{remark}
\newtheorem{remark}{Remark}

\journal{Elsevier}

\newlength{\highlightswidth}


\begin{document}

\begin{frontmatter}

\title{Nonlinear homogenization of periodic unit cells composed of mixed continuum and structural elements with variationally consistent boundary-member reduction}

\author[]{Jonathan Stollberg\corref{cor1}}
\ead{jonathan.stollberg@tu-darmstadt.de}
\cortext[cor1]{Corresponding author}

\author[]{Dominik Schillinger}
\ead{dominik.schillinger@tu-darmstadt.de}

\affiliation{organization={Institute for Mechanics, Computational Mechanics
Group, Technical University of Darmstadt}, city={Darmstadt}, postcode={64287}, country={Germany}}

\begin{abstract}
Periodic architected materials may be represented by representative volume elements (RVEs) combining continuum, truss, beam, and shell constituents. Their mixed translational and rotational kinematics complicate a consistent scale transition to a macroscale Cauchy continuum. We establish a nonlinear first-order computational homogenization framework for such mixed-dimensional RVEs. An auxiliary finite-strain Cosserat continuum serves solely as a power-equivalent carrier of their virtual power and generalized boundary kinematics. Its uniform rotation is identified with the macroscopic polar rotation and its wryness vanishes, so the effective medium remains of Cauchy type. Under periodic translational and rotational constraints, the Hill--Mandel condition yields a common nodal-force virial for the effective first Piola--Kirchhoff stress, independent of element type. For lower-dimensional members on periodically identified cell boundaries, fractional ownership of internal virtual work is shown to be variationally equivalent to a topology-preserving reduction retaining one representative of each set of periodic element copies. Because this equivalence covers the complete element virtual work and consistent linearization, it applies to geometrically and materially nonlinear and path-dependent constituents without rescaling section properties, constitutive parameters, or internal variables. Numerical examples recover the analytical small-strain response and cubic symmetry of a mixed continuum--shell--beam cell, verify the reduction and discrete power balance, and demonstrate finite-rotation buckling of a hyperelastic Octet lattice and history-dependent yield-surface evolution of an elastoplastic sandwich cell. The open-source implementation \texttt{abqhom} couples the framework to Abaqus.
\end{abstract}

% \begingroup
% \setlength{\textwidth}{\highlightswidth}
% \setlength{\columnwidth}{\highlightswidth}
% \setlength{\linewidth}{\highlightswidth}
% \setlength{\medskipamount}{6pt}
% \begin{highlights}
% \setlength{\topsep}{0.5\baselineskip}
% \setlength{\itemsep}{0.35\baselineskip}
% \setlength{\parsep}{0pt}
% \setlength{\parskip}{0pt}
% \setlength{\partopsep}{0pt}
% \item One nonlinear framework homogenizes mixed continuum and structural RVEs.
% \item An auxiliary Cosserat power carrier links mixed RVEs to a Cauchy macroscale.
% \item Boundary-member ownership is variationally equivalent to periodic element reduction.
% \item The reduction supports nonlinear and path-dependent constitutive behavior.
% \item Open-source library automates mixed-dimensional RVE analyses with Gmsh and Abaqus.
% \end{highlights}
% \endgroup

\begin{keyword}
Computational homogenization \sep Periodic unit cells \sep Mixed-dimensional
finite elements \sep Lattice structures \sep Finite-strain mechanics \sep
Hill--Mandel condition
\end{keyword}

\end{frontmatter}

% ========================================================================================
\section{Introduction}\label{sec:introduction}%
Architected cellular materials such as lattices, foams, and triply periodic minimal surface (TPMS) solids derive their distinctive macroscopic behavior from the interplay between the base material and the geometry of their unit cell, achieving combinations of stiffness and strength, and unconventional responses such as negative Poisson's ratio, that are difficult to realize with conventional homogeneous solids \citep{gibson_1997, fleck_2010, bertoldi_2017}. Additive manufacturing realizes ever more complex cells whose slender members carry forces together with bending and twisting moments and deform well into the finite-strain regime, where they buckle and rotate \citep{surjadi_2019}. As resolving every member at component scale is prohibitive, computational homogenization, which replaces the microstructure with an energetically equivalent effective continuum, is the natural remedy.

Homogenization rests on the asymptotic theory of periodic media \citep{bensoussan_1978, sanchezpalencia_1980} and on the averaging and macrohomogeneity principles of Hill \citep{hill_1963, hill_1972} and Mandel \citep{mandel_1972}, which relate macroscopic measures to averages over microscopic representative volume elements (RVEs) and enforce energetic consistency between the scales. These principles underlie the variational unit-cell formulation of \citet{suquet_1987} and its finite element realization, through which the effective RVE response is obtained from a microscopic boundary-value problem \citep{terada_1995, miehe_1999, terada_2001}. Embedding this problem into a nested macro--micro finite element architecture leads to the approach now termed FE\textsuperscript{2} \citep{smit_1998, feyel_2000, kouznetsova_2001}. Subsequent work addressed RVE convergence and boundary conditions \citep{terada_2000}, representative cell size \citep{kanit_2003, gitman_2007}, consistent macroscopic tangents \citep{miehe_2003, temizer_2008}, and generalized-continuum macromodels \citep{feyel_2003, kouznetsova_2004}. Broader overviews are given in \cite{nematnasser_1999, geers_2010}. Most first-order formulations resolve the microstructure as a Cauchy continuum with translational degrees of freedom. This suits heterogeneous composites, but not discrete beam-, plate-, or shell-type microstructures that carry force and moment resultants and possess explicit rotations.

A parallel line addresses cellular solids. Analytical scaling laws connect relative density and stiffness and distinguish bending- from stretching-dominated architectures \citep{gibson_1997, deshpande_2001}, but are restricted to idealized topologies and do not describe the nonlinear response of general cells. Unit-cell-based asymptotic and numerical schemes subsequently enabled the effective elastic properties and strength limits of more general beam and truss cells to be evaluated \citep{vigliotti_2012a, vigliotti_2012b, arabnejad_2013}. More recent work extended structural-element RVEs to finite rotations, localization, and nonlinear buckling \citep{vigliotti_2014, glaesener_2019, glaesener_2020, jamshidian_2020, herrnbock_2022}, to buckling-aware two-scale optimization and elastoplastic response \citep{hubner_2023, danesh_2025}, and to linear-elastic shell-lattice homogenization \citep{shojaee_2024}. Boundary-member duplication has meanwhile been avoided through reduced fundamental cells in finite-strain beam-lattice studies \citep{gartner_2021, fernandez_2022} or through periodic rotational constraints combined with fractional cross sections \citep{schwahofer_2023}. These developments establish structural-element RVEs as an efficient alternative to fully resolved solids, but remain segmented by member class, constitutive regime, or topology.

Independent member rotations also raise the question whether the effective medium should be generalized rather than Cauchy. Such continua originate with the oriented-body model of the Cosserat brothers \citep{cosserat_1909} and its development into couple-stress and micropolar elasticity \citep{toupin_1962, mindlin_1964, eringen_1966}. The appropriate description depends on the heterogeneity size relative to an intrinsic length \citep{forest_1998, forest_2001}. Micropolar effective continua have been constructed from beam lattices, cellular materials, and discrete models with rotational degrees of freedom \citep{dosreis_2012, bacigalupo_2014, rezakhani_2016, gad_2021}, while second-order beam-network homogenization retains strain-gradient effects \citep{weeger_2021, ye_2024}. Closer to the present construction, a Cosserat-equivalent RVE has been used to transfer boundary conditions and averaged measures between a discrete particle assembly and a Cosserat macroscale \citep{li_2010}, while virtual-work equivalence provides macroscopic stress and couple-stress measures for finite heterogeneous Cosserat continua and particle systems with independent translations and rotations \citep{elias_2025}. In both cases, the Cosserat description is assumed or effective rather than an auxiliary representation for a Cauchy scale transition. Discrete rotational models may themselves homogenize to either a Cauchy or a Cosserat continuum depending on their bending-stiffness scaling \citep{elias_2026}, and finite-strain micropolar-to-micropolar homogenization has also been established \citep{thiesen_2026}. Although generalized macromodels capture size effects and moment transmission, microscopic rotations alone do not imply a Cosserat macroscale \citep{elias_2026}. Here they remain in the RVE while the effective medium is deliberately restricted to a first-order Cauchy continuum compatible with conventional solvers.

Open-source implementations address continuum RVEs within a commercial finite element environment \citep{omairey_2019}, solid-, beam-, and shell-discretized cells including a compatible mixed beam--shell configuration \citep{liu_2026}, and standalone workflows for truss-, plate-, and shell-based lattices \citep{jin_2026}. Their demonstrated scope remains restricted to small-deformation linear elasticity, leaving open a common finite-strain transition for compatible mixtures of continuum and structural constituents. Treatments of lower-dimensional boundary members likewise rely on cells reduced a priori or topology-specific section corrections rather than a general variational relation between the complete cell and a topology-preserving reduced nonlinear RVE.

Against this background, this work makes two principal contributions. First, it develops a common first-order finite-strain scale transition for periodic RVEs combining compatible continuum, truss, beam, and shell constituents. An auxiliary Cosserat continuum represents their total virtual power and generalized boundary kinematics without becoming the effective medium. Identifying the uniform rotation factor of this representation with the macroscopic polar rotation and setting the associated macroscopic wryness to zero retain microscopic rotations and moments in the RVE equilibrium while restricting the macroscale to a Cauchy continuum. Second, periodic ownership factors of lower-dimensional boundary constituents are introduced through the internal virtual work and shown to be variationally equivalent to a topology-preserving reduction retaining one representative of each periodic element set. Here, equivalence between the ownership-weighted complete cell and the topology-preserving reduced model is established at the level of the complete element virtual work and its consistent linearization. The reduced problem therefore accommodates geometric and material nonlinearity and path-dependent behavior without rescaling sections, material parameters, or history variables. These contributions are implemented in the open-source library \texttt{abqhom}, coupling the mesh generator Gmsh \citep{geuzaine_2009} to the commercial finite element solver Abaqus.

The remainder of this paper is organized as follows. Section~\ref{sec:cosserat} recalls the finite-strain Cosserat kinematics and virtual-work statement. Section~\ref{sec:homogenization} develops the scale transition, Hill--Mandel condition, and effective stress measures. Section~\ref{sec:software} gives the discrete procedure, Section~\ref{sec:results} validates and applies the method, and Section~\ref{sec:conclusion} concludes.

% ========================================================================================
\section{Finite-strain Cosserat continua}\label{sec:cosserat}%
This section summarizes the finite-strain Cosserat kinematics and virtual-work framework. Section~\ref{sec:kinematics} introduces the independent translational and rotational fields and their strain measures, while Section~\ref{sec:virtual_work} states the corresponding virtual-work principle, stresses, balance laws, and natural boundary quantities.

\begin{figure*}[htbp]
    \centering
    \includegraphics[width=0.8\textwidth]{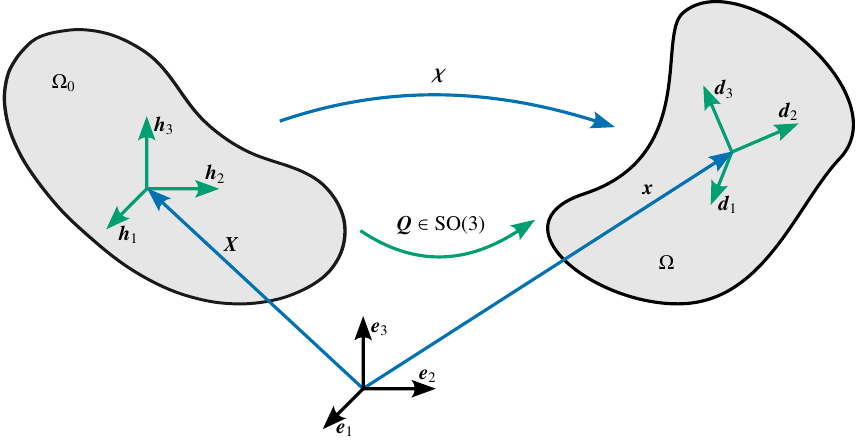}
    \caption{Finite-strain Cosserat kinematics: the deformation map $\boldsymbol{\chi}$ maps $\boldsymbol{X} \in \Omega_{0}$ to $\boldsymbol{x} \in \Omega$, while the independent microrotation $\boldsymbol{Q} \in \mathrm{SO}(3)$ maps the reference director frame $\{\boldsymbol{h}_{a}\}_{a = 1}^{3}$ to the current frame $\{\boldsymbol{d}_{a}\}_{a = 1}^{3}$ relative to the fixed Cartesian basis $\{\boldsymbol{e}_{i}\}_{i = 1}^{3}$.}
    \label{fig:cosserat_kinematics}%
\end{figure*}

% ----------------------------------------------------------------------------------------
\subsection{Micropolar kinematics}\label{sec:kinematics}%
Let $\Omega_{0} \subset \mathbb{R}^{3}$ be the Lagrangian reference domain of a continuum $\mathcal{C}$ with measure $V \coloneqq \lvert \Omega_{0} \rvert$, boundary $\partial \Omega_{0}$, and outward unit normal $\boldsymbol{N}$. Infinitesimal volume and surface elements are denoted $\mathrm{d} V$ and $\mathrm{d} A$, respectively. A material point at the reference position $\boldsymbol{X} \in \Omega_{0}$ is described in the current configuration $\Omega$ by a smooth, invertible deformation map $\boldsymbol{\chi} : \Omega_{0} \to \Omega$ and by an independent orthonormal director frame $\{\boldsymbol{d}_{a}\}_{a = 1}^{3}$. Its spatial position is $\boldsymbol{x} \coloneqq \boldsymbol{\chi}(\boldsymbol{X}) \in \Omega$. Relative to a fixed orthonormal reference director frame $\{\boldsymbol{h}_{a}\}_{a = 1}^{3}$, the finite Cosserat motion is written as
\begin{subequations}
    \begin{align}
        \boldsymbol{x}(\boldsymbol{X}) &= \boldsymbol{\chi}(\boldsymbol{X}) = \boldsymbol{X} + \boldsymbol{u}(\boldsymbol{X}) \,,\\
        \boldsymbol{d}_{a}(\boldsymbol{X}) &= \boldsymbol{Q}(\boldsymbol{X}) \boldsymbol{h}_{a} \,,
    \end{align}
\end{subequations}
where $\boldsymbol{u} = \boldsymbol{u}(\boldsymbol{X})$ is the translational displacement field and 
\begin{align}
    \boldsymbol{Q}(\boldsymbol{X}) &\coloneqq \sum_{a = 1}^{3} \boldsymbol{d}_{a} \otimes \boldsymbol{h}_{a} \in \mathrm{SO}(3) \,,
\end{align}
denotes the independent microrotation tensor of the three-dimensional rotation group
\begin{align}
    \mathrm{SO}(3) \coloneqq \bigl\{ \boldsymbol{Q} \mid \boldsymbol{Q}^{\mathrm{T}} \boldsymbol{Q} = \boldsymbol{I},\ \det \boldsymbol{Q} = 1 \bigr\} \,.
\end{align}
The undeformed reference state is represented by $\boldsymbol{x} = \boldsymbol{X}$ and $\boldsymbol{Q} = \boldsymbol{I}$, so that the current and reference director frames initially coincide. The symbol $\boldsymbol{I}$ represents the second-order identity tensor. Thus, the continuum assigns a nonstandard set of independent kinematic degrees of freedom to each material point, namely three translations and three rotations. This is the essential kinematic distinction between a Cosserat and a classical Cauchy continuum \citep{cosserat_1909, eringen_1966, eringen_1964}. Figure~\ref{fig:cosserat_kinematics} summarizes this distinction. We emphasize that the two-dimensional setup follows by the corresponding planar reduction.

The corresponding finite strain measures we consider in this work are the deformation gradient, the relative deformation tensor, and the wryness tensor \citep{pietraszkiewicz_2009},
\begin{subequations}
    \begin{align}
        \boldsymbol{F} &\coloneqq \operatorname{Grad} \boldsymbol{x} \,,\\
        \boldsymbol{E} &\coloneqq \boldsymbol{Q}^{\mathrm{T}} \boldsymbol{F} - \boldsymbol{I} \,,\\
        \boldsymbol{K} &\coloneqq \sum_{i = 1}^{3} \operatorname{axl} \left( \boldsymbol{Q}^{\mathrm{T}} \frac{\partial \boldsymbol{Q}}{\partial X_{i}} \right) \otimes \boldsymbol{e}_{i} \,,
    \end{align}
    \label{eq:cosserat_strain_measures}%
\end{subequations}
respectively, where $\{\boldsymbol{e}_{i}\}_{i = 1}^{3}$ is a fixed Cartesian basis in the reference configuration and $X_{i} \coloneqq \boldsymbol{X} \cdot \boldsymbol{e}_{i}$. The gradient operator with respect to reference coordinates is denoted $\operatorname{Grad}$. Only orientation-preserving motions are admitted, so that $\det \boldsymbol{F} > 0$ throughout $\Omega_{0}$. Moreover, for any vector $\boldsymbol{a} \in \mathbb{R}^{3}$, its associated skew-symmetric tensor $\widehat{\boldsymbol{a}} \in \mathbb{R}^{3 \times 3}$ is defined by
\begin{align}
    \widehat{\boldsymbol{a}} \boldsymbol{b} = \boldsymbol{a} \times \boldsymbol{b} \qquad \forall\,\boldsymbol{b} \in \mathbb{R}^{3} \,,
    \label{eq:hat_operator_definition}%
\end{align}
and $\operatorname{axl}$ denotes the inverse mapping, $\operatorname{axl}(\widehat{\boldsymbol{a}}) = \boldsymbol{a}$. Note that both $\boldsymbol{E}$ and $\boldsymbol{K}$ are generally non-symmetric and objective.

% ----------------------------------------------------------------------------------------
\subsection{Principle of virtual work}\label{sec:virtual_work}%
For an admissible virtual motion, define the spatial virtual-spin tensor $\widehat{\boldsymbol{\eta}}$ by
\begin{align}
    \widehat{\boldsymbol{\eta}} \coloneqq \delta \boldsymbol{Q} \boldsymbol{Q}^{\mathrm{T}} \,,
    \label{eq:virtual_spin_definition}%
\end{align}
and let $\boldsymbol{\eta} \coloneqq \operatorname{axl}(\widehat{\boldsymbol{\eta}})$ be the associated virtual-rotation vector. Direct variation of \eqref{eq:cosserat_strain_measures} gives the exact kinematic identities
\begin{subequations}
    \label{eq:cosserat_variations}%
    \begin{align}
        \delta \boldsymbol{F} &= \operatorname{Grad} \delta \boldsymbol{u} \,,\\
        \delta \boldsymbol{E} &= \boldsymbol{Q}^{\mathrm{T}}\bigl( \delta \boldsymbol{F} - \widehat{\boldsymbol{\eta}} \boldsymbol{F} \bigr) \,,\\
        \delta \boldsymbol{K} &= \boldsymbol{Q}^{\mathrm{T}} \operatorname{Grad} \boldsymbol{\eta} \,.
    \end{align}
\end{subequations}

We introduce the material force-stress tensor $\boldsymbol{\Sigma}$ and the material couple-stress tensor $\boldsymbol{M}$ through the internal virtual work of the Cosserat continuum
\begin{align}
    \delta W^{\mathrm{int}}_{\mathcal{C}} \coloneqq \int_{\Omega_{0}} \bigl( \boldsymbol{\Sigma} : \delta \boldsymbol{E} + \boldsymbol{M} : \delta \boldsymbol{K} \bigr) \,\mathrm{d} V \,,
    \label{eq:internal_virtual_work}%
\end{align}
such that $\boldsymbol{\Sigma}$ is work-conjugate to the relative deformation tensor $\boldsymbol{E}$, whereas $\boldsymbol{M}$ is work-conjugate to the wryness $\boldsymbol{K}$. Their exact formulation depends on the constitutive specification. For example, in the particular case of a hyperelastic Cosserat material with Helmholtz free-energy density per unit reference volume $\Psi = \Psi(\boldsymbol{E}, \boldsymbol{K})$, the stresses specialize to
\begin{align}
    \boldsymbol{\Sigma} = \frac{\partial \Psi}{\partial \boldsymbol{E}} \,,\qquad \boldsymbol{M} = \frac{\partial \Psi}{\partial \boldsymbol{K}} \,.
\end{align}

\begin{figure*}[htbp]
    \centering
    \includegraphics[width=0.8\textwidth]{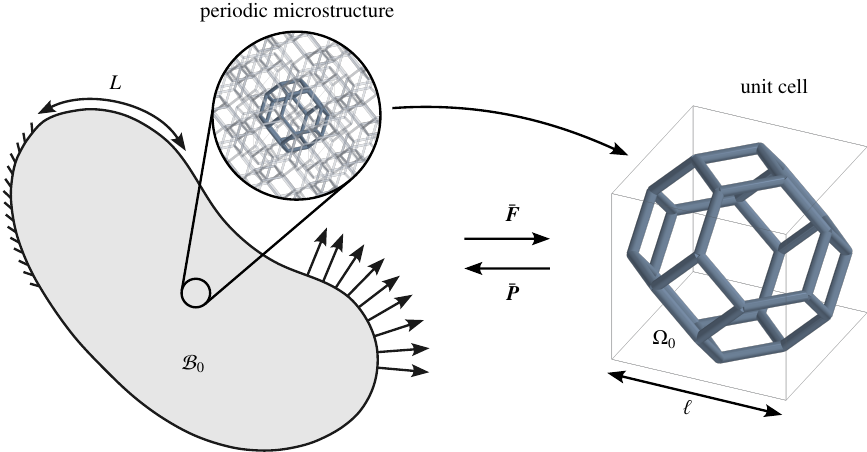}
    \caption{Multiscale setting of the homogenization framework. Each material point of the macroscopic body $\mathcal{B}_{0}$, with characteristic length $L$, is associated with a periodic RVE $\Omega_{0}$, illustrated by a Kelvin beam lattice, with characteristic length $\ell \ll L$. Homogenization maps the prescribed $\bar{\boldsymbol{F}}$ to the effective stress $\bar{\boldsymbol{P}}$.}
    \label{fig:rve_extraction}%
\end{figure*}

To express \eqref{eq:internal_virtual_work} in terms of the spatial virtual translations and rotations, introduce the first Piola--Kirchhoff force-stress tensor $\boldsymbol{P}$, the internal moment vector $\boldsymbol{r}$ generated by its skew part, and the couple-stress tensor $\boldsymbol{\mu}$ as
\begin{subequations}
    \begin{align}
        \boldsymbol{P} &\coloneqq \boldsymbol{Q} \boldsymbol{\Sigma} \,,\\
        \boldsymbol{r} &\coloneqq 2 \operatorname{axl} \bigl( \operatorname{skew}( \boldsymbol{P} \boldsymbol{F}^{\mathrm{T}} ) \bigr) \,,\label{eq:internal_moment_definition}%
        \\
        \boldsymbol{\mu} &\coloneqq \boldsymbol{Q} \boldsymbol{M} \,,
    \end{align}
\end{subequations}
respectively, where $\operatorname{skew}( \boldsymbol{A} ) \coloneqq \tfrac{1}{2}( \boldsymbol{A} - \boldsymbol{A}^{\mathrm{T}} )$ for any $\boldsymbol{A} \in \mathbb{R}^{3 \times 3}$. With the axial-vector convention used above, the identities
\begin{subequations}
    \begin{align}
        \boldsymbol{\Sigma} : \bigl( \boldsymbol{Q}^{\mathrm{T}} \widehat{\boldsymbol{\eta}} \boldsymbol{F} \bigr) &= ( \boldsymbol{P} \boldsymbol{F}^{\mathrm{T}} ) : \widehat{\boldsymbol{\eta}} \,,\\
        ( \boldsymbol{P} \boldsymbol{F}^{\mathrm{T}} ) : \widehat{\boldsymbol{\eta}} &= \boldsymbol{r} \cdot \boldsymbol{\eta}
    \end{align}
\end{subequations}
relate the force-stress contribution associated with the virtual spin to the virtual-rotation vector. Substitution of the kinematic variations from \eqref{eq:cosserat_variations} into the internal virtual work, followed by use of these identities, gives
\begin{align}
     \delta W^{\mathrm{int}}_{\mathcal{C}} = \int_{\Omega_{0}} \bigl( \boldsymbol{P} : \operatorname{Grad} \delta \boldsymbol{u} - \boldsymbol{r} \cdot \boldsymbol{\eta} + \boldsymbol{\mu} : \operatorname{Grad} \boldsymbol{\eta} \bigr) \,\mathrm{d} V \,.
\end{align}
Assuming the absence of body forces and body couples, let $\boldsymbol{t}$ and $\boldsymbol{c}$ denote the referential boundary force traction and couple traction, respectively. Their external virtual work is
\begin{align}
    \delta W^{\mathrm{ext}}_{\mathcal{C}} \coloneqq \int_{\partial \Omega_{0}} \bigl( \boldsymbol{t} \cdot \delta \boldsymbol{u} + \boldsymbol{c} \cdot \boldsymbol{\eta} \bigr) \,\mathrm{d} A \,,
    \label{eq:cosserat_external_virtual_work}%
\end{align}
with the virtual fields vanishing for parts of the boundary on which the corresponding kinematic quantities are prescribed. The quasi-static principle of virtual work therefore requires
\begin{align}
    \delta W^{\mathrm{int}}_{\mathcal{C}} = \delta W^{\mathrm{ext}}_{\mathcal{C}} \,,
    \label{eq:cosserat_virtual_work_principle}%
\end{align}
for every admissible virtual motion. Applying the divergence theorem to the terms containing $\operatorname{Grad} \delta \boldsymbol{u}$ and $\operatorname{Grad} \boldsymbol{\eta}$ in the internal virtual work, and invoking \eqref{eq:cosserat_virtual_work_principle} as well as the arbitrariness of the admissible virtual fields, yields the local quasi-static equilibrium conditions
\begin{subequations}
    \begin{align}
        \operatorname{Div} \boldsymbol{P} &= \boldsymbol{0} \,,\label{eq:force_balance}%
        \\
        \operatorname{Div} \boldsymbol{\mu} + \boldsymbol{r} &= \boldsymbol{0} \,.\label{eq:couple_balance}%
    \end{align}
\end{subequations}
Here $\operatorname{Div}$ denotes the divergence with respect to the reference coordinates. The boundary terms identify the corresponding natural tractions as $\boldsymbol{t} = \boldsymbol{P} \boldsymbol{N}$ and $\boldsymbol{c} = \boldsymbol{\mu} \boldsymbol{N}$. This boundary-power representation provides the common continuum description used in the following section to relate the mixed-dimensional structural RVE to the Cauchy macroscale body.

% ========================================================================================
\section{Computational homogenization using an auxiliary Cosserat representation of the microstructure}\label{sec:homogenization}%
Computational homogenization determines the effective response of a heterogeneous material from a boundary-value problem on a representative volume element of the microstructure \citep{suquet_1987, kouznetsova_2001, miehe_2003, geers_2010}. To this end, consider a macroscopic body $\mathcal{B}_{0}$ with characteristic length $L$ and an associated microstructure with characteristic length $\ell$. A separation of scales is assumed:
\begin{align}
    \frac{\ell}{L} \ll 1 \,.
\end{align}
In this work, we restrict our attention to periodic microstructures, so that the RVE is a repeating unit cell, as illustrated in Figure~\ref{fig:rve_extraction}. The macroscale body is taken to be a classical first-order Cauchy continuum whose kinematics are described solely by the macroscopic deformation gradient $\bar{\boldsymbol{F}} = \bar{\boldsymbol{R}}\bar{\boldsymbol{U}}$, where $\bar{\boldsymbol{R}}$ and $\bar{\boldsymbol{U}}$ are its polar rotation and right stretch tensors, respectively, and $\det\bar{\boldsymbol{F}}>0$. It therefore possesses neither an independent Cosserat rotation field nor a curvature--couple-stress power pair. Within the local first-order Cosserat representation developed below, the uniform rotation factor $\bar{\boldsymbol{R}}$ is constant over the RVE and the associated macroscopic wryness is consequently zero.

This section develops the homogenization framework for RVEs built from continuum, truss, beam, or shell formulations. Section~\ref{subsec:structural_rve} defines their mixed-dimensional material support and ownership-corrected internal virtual work. Section~\ref{subsec:cosserat_equivalence} introduces the power-equivalent Cosserat representation, Section~\ref{subsec:hillmandel} derives the admissible boundary conditions and effective stress from the Hill--Mandel condition, and Section~\ref{subsec:smallstrain} gives the geometrically linear specialization.

% ----------------------------------------------------------------------------------------
\subsection{Periodic structural representative volume element}\label{subsec:structural_rve}%
The present work is restricted to cuboidal periodic cells with faces aligned with the coordinate axes associated with the Cartesian basis $\{\boldsymbol{e}_{i}\}_{i=1}^{3}$. Let $\Omega_{0}$ now denote such a reference cell, with volume $V \coloneqq \lvert \Omega_{0} \rvert$ and boundary $\partial \Omega_{0}$. Up to their lower-dimensional intersections, the boundary is the union of three pairs of opposite faces $\Gamma_{i}^{\pm}$, whose outward unit normals satisfy $\boldsymbol{N}_{i}^{+} = -\boldsymbol{N}_{i}^{-}$. Each point $\boldsymbol{X}^{-}\in\Gamma_{i}^{-}$ has a matching point $\boldsymbol{X}^{+}\in\Gamma_{i}^{+}$ related by the constant cell translation $\Delta\boldsymbol{X}^{i}$,
\begin{subequations}
	\begin{align}
    	\partial\Omega_{0} &= \bigcup_{i=1}^{3}( \Gamma_{i}^{+}\cup\Gamma_{i}^{-} )\,,\\
    	\Delta\boldsymbol{X}^{i} &= \boldsymbol{X}^{+}-\boldsymbol{X}^{-} \,.
	\end{align}
\end{subequations}
The translation vectors generate the infinite lattice
\begin{align}
    \mathcal{L} \coloneqq \left\{ \sum_{i=1}^{3} n_{i} \Delta \boldsymbol{X}^{i} \;\middle\vert\; n_{i} \in\mathbb{Z} \right \}\,,
\end{align}
under which the material support, its constitutive properties, and its initial material state are periodic. The volume $V$ of the enclosing cell, rather than the volume of its material constituents, is used for the scale transition and therefore includes any voids.

\begin{figure}[htbp]
    \centering
    \includegraphics[width=0.95\columnwidth]{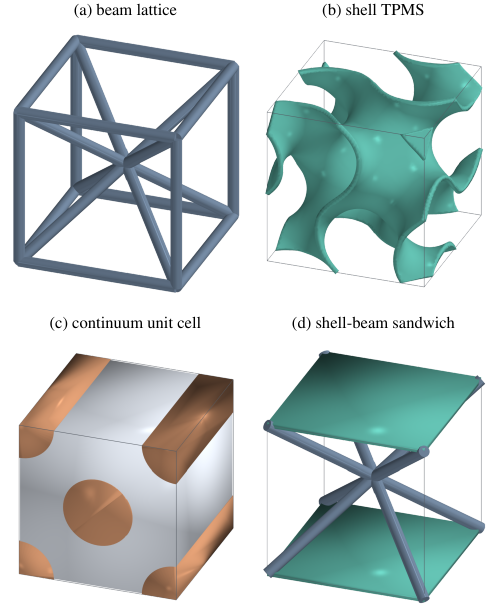}
    \caption{Examples of compatible constituent combinations within the cuboidal structural RVE $\mathcal{S}$: (a) beam lattice, (b) shell-based triply periodic minimal surface, (c) fiber-reinforced continuum, and (d) shell--beam sandwich.}
    \label{fig:unit_cell_variety}%
\end{figure}

We define the set of constituent formulations
\begin{align}
    \mathfrak{M} \coloneqq \{ \mathrm{Cauchy}, \mathrm{truss}, \mathrm{beam}, \mathrm{shell} \} \,,
\end{align}
indexed by $m\in\mathfrak{M}$. The reference material support of the structural RVE $\mathcal{S}$ is
\begin{align}
    \mathcal{M}_{0} \coloneqq \bigcup_{m \in \mathfrak{M}} \mathcal{M}_{0}^{m} \,,
\end{align}
where each $\mathcal{M}_{0}^{m}$ may be a disconnected union. In particular, $\mathcal{M}_{0}^{\mathrm{Cauchy}}$ comprises three-dimensional Cauchy-continuum regions, $\mathcal{M}_{0}^{\mathrm{truss}}$ and $\mathcal{M}_{0}^{\mathrm{beam}}$ comprise the respective one-dimensional centerlines, and $\mathcal{M}_{0}^{\mathrm{shell}}$ comprises two-dimensional shell midsurfaces. Integration over each support employs its intrinsic measure,
\begin{align}
    \mathrm{d} \mu_{m} \coloneqq
    \begin{cases}
        \mathrm{d} V \,, & m = \mathrm{Cauchy} \,,\\
        \mathrm{d} s \,, & m \in \{ \mathrm{truss}, \mathrm{beam} \} \,,\\
        \mathrm{d} A \,, & m = \mathrm{shell} \,.
    \end{cases}
\end{align}
This notation allows each constituent formulation to retain its natural one-, two-, or three-dimensional virtual-work density and corresponding integration measure. Because each constituent is treated in its own natural setting, these supports may be combined in compatible arrangements, as illustrated in Figure~\ref{fig:unit_cell_variety}. Where different constituent formulations meet, their kinematics are coupled such that the resulting interface power is consistent. Conforming beam--beam and beam--shell connections are treated as kinematically rigid through shared nodal translations and rotations. Their finite joint geometry and associated local stiffening are not resolved, but may be incorporated through geometry-based enrichments \citep{weeger_2019}. Under the periodic identification of matching boundaries, the assembled support is required to form a kinematically connected structural model, which is considered in quasi-static equilibrium. Its admissible virtual motion comprises translations on all supports and, additionally, independent cross-section rotations for beams and director rotations for shells. For the latter formulations, the chosen objective finite-rotation theory supplies the precise strain measures and work-conjugate resultants: generalized forces are conjugate to extension and shear strain variations, while generalized moments are conjugate to bending, torsion, or curvature variations \citep{simo_1985, simo_1989}. The contribution of each constituent is evaluated on its own support, and the total structural power is obtained by summing these contributions. Kinematic coupling transmits forces and moments between constituents without altering their intrinsic integration measures. The same assembly therefore accommodates homogeneous and mixed-dimensional cells without changing the subsequent homogenization argument. On the RVE boundary, the generalized forces and moments are power-conjugate to the admissible virtual translations and spatial virtual rotations. This common work-conjugate description underlies the subsequent scale transition.

\begin{figure*}[htbp]
    \centering
    \includegraphics[width=0.95\textwidth]{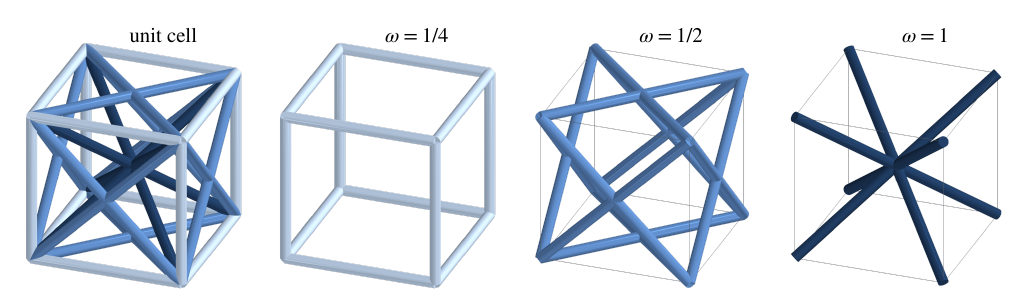}%
    \caption{Geometric ownership of boundary-supported constituents in a lattice cell combining simple-cubic, face-centered-cubic, and body-centered-cubic struts. From left: complete cell and its edge ($\omega = 1/4$), face ($\omega = 1/2$), and interior ($\omega = 1$) members.}
    \label{fig:sharing}%
\end{figure*}

Special care is required for material support contained in the periodically identified cell boundary. A constituent portion on a face is shared by two translated cells in the infinite lattice $\mathcal{L}$, while a constituent portion on the intersection of two faces is shared by four. To assign the internal virtual work of the infinite periodic structure to one cell, define for $\boldsymbol{X} \in \Omega_{0}$
\begin{align}
    \mathcal{F}_{\boldsymbol{X}} \coloneqq \left\{ \Gamma_{i}^{\varsigma} \;\middle\vert\; i \in \{ 1, 2, 3 \},\; \varsigma \in \{ +, - \},\; \boldsymbol{X} \in \Gamma_{i}^{\varsigma} \right\}
    \label{eq:boundary_face_set}%
\end{align}
as the set of cell faces containing $\boldsymbol{X}$. Its cardinality determines the ownership factor
\begin{align}
    \omega(\boldsymbol{X}) &\coloneqq 2^{-\lvert \mathcal{F}_{\boldsymbol{X}} \rvert}\,.
\end{align}
Hence $\omega = 1$ in the cell interior, $\omega = 1/2$ on a face, and $\omega = 1/4$ on an edge. The value $\omega = 1/8$ at a corner is immaterial for the present continuum, beam, truss, and shell constituents because isolated points have zero intrinsic measure. More generally, lower-dimensional intersections of a constituent support with the cell boundary do not alter its virtual work: a beam that merely terminates at or crosses a face retains unit ownership almost everywhere along its centerline, whereas a beam centerline contained in that face carries half ownership. The same distinction applies to shell midsurfaces and continuum regions with respect to their intrinsic measures. Figure~\ref{fig:sharing} illustrates this ownership construction for line-like constituents.

Let $\delta w_{m}^{\mathrm{int}}$ denote the internal virtual-work density supplied by formulation $m$. The internal virtual work assigned to the structural RVE is then
\begin{align}
    \delta W_{\mathcal{S}}^{\mathrm{int}} \coloneqq \sum_{m \in \mathfrak{M}} \int_{\mathcal{M}_{0}^{m}} \omega( \boldsymbol{X} ) \delta w_{m}^{\mathrm{int}} \,\mathrm{d} \mu_{m} \,.
    \label{eq:ownership_corrected_internal_work}%
\end{align}
An absent constituent formulation is represented by an empty support and therefore contributes zero. Section~\ref{sec:software} develops a variationally equivalent discrete realization that avoids fractional modification of the element formulations.

% ----------------------------------------------------------------------------------------
\subsection{Power-equivalent Cosserat representation}\label{subsec:cosserat_equivalence}%
The mixed-dimensional material support $\mathcal{M}_{0}$ of the structural RVE $\mathcal{S}$ does not possess a single common three-dimensional pointwise stress field on $\Omega_{0}$. To formulate a common scale transition for its different constituent formulations, an auxiliary continuum $\mathcal{C}$ is introduced as a continuous carrier of the same total virtual power. We postulate that $\mathcal{C}$ exists as a sufficiently regular, standard Cosserat continuum with the kinematic fields $\boldsymbol{u}$ and $\boldsymbol{Q}$, the stress fields $\boldsymbol{P}$ and $\boldsymbol{\mu}$, and the virtual-work structure introduced in Section~\ref{sec:cosserat}. It is defined on the entire reference cell $\Omega_{0}$ and therefore has the same outer boundary $\partial \Omega_{0}$ and reference volume $V = \lvert \Omega_{0} \rvert$ as the structural RVE.

On the portions of the material support of $\mathcal{S}$ located on the cell boundary $\partial\Omega_{0}$, the translational kinematics of its Cauchy regions, beam and truss centerlines, and shell midsurfaces are identified with the corresponding boundary values of $\boldsymbol{u}$. Likewise, the finite cross-section rotations of beams and the finite director rotations of shells are identified with the corresponding boundary values of $\boldsymbol{Q}$. The associated admissible structural virtual translations and rotations are identified with the corresponding boundary values of $\delta\boldsymbol{u}$ and $\boldsymbol{\eta}$. The structural and Cosserat boundary virtual motions are thus not independent. For each admissible structural virtual motion, we assume the existence of a sufficiently regular auxiliary virtual motion satisfying these boundary identifications and impose power equivalence for the resulting pair:
\begin{align}
    \delta W^{\mathrm{int}}_{\mathcal{S}} = \delta W^{\mathrm{int}}_{\mathcal{C}} \,.
    \label{eq:cosserat_power_equivalence}%
\end{align}
For paired admissible virtual motions about an equilibrated state, this equality and the principle of virtual work \eqref{eq:cosserat_virtual_work_principle} give
\begin{align}
    \begin{split}
        \delta W^{\mathrm{int}}_{\mathcal{S}} &= \delta W^{\mathrm{int}}_{\mathcal{C}} = \delta W^{\mathrm{ext}}_{\mathcal{C}} \\
        &= \int_{\partial\Omega_{0}} \bigl( \boldsymbol{t}\cdot\delta\boldsymbol{u} + \boldsymbol{c}\cdot\boldsymbol{\eta} \bigr) \,\mathrm{d}A\,.
    \end{split}
\end{align}
Hence the boundary-power functional generated by the auxiliary tractions equals the boundary virtual work of the generalized forces and moments of $\mathcal{S}$. Related formulations link discrete particle assemblies to Cosserat-equivalent RVEs \citep{li_2010} and use virtual-work equivalence to derive macroscopic stress and couple-stress measures from finite heterogeneous Cosserat continua and particle systems \citep{elias_2025}. In the present setting, $\mathcal{C}$ instead provides an auxiliary, power-equivalent representation of the total virtual power and generalized boundary kinematics of the mixed-dimensional RVE.

% ----------------------------------------------------------------------------------------
\subsection{Hill--Mandel condition, admissible boundary conditions, and effective stress}\label{subsec:hillmandel}%
The microscopic placement and displacement are decomposed into a macroscopic affine part and a displacement fluctuation \citep{miehe_2003},
\begin{subequations}
    \label{eq:first_order_localization}%
    \begin{align}
        \boldsymbol{x}(\boldsymbol{X}) &= \bar{\boldsymbol{F}} \boldsymbol{X} + \tilde{\boldsymbol{u}}(\boldsymbol{X}) \,,\\
        \boldsymbol{u}(\boldsymbol{X}) &= ( \bar{\boldsymbol{F}} - \boldsymbol{I} ) \boldsymbol{X} + \tilde{\boldsymbol{u}}(\boldsymbol{X}) \,.
    \end{align}
\end{subequations}
The displacement fluctuation remains undetermined up to an additive rigid translation, which is fixed when posing the microscopic boundary-value problem. Taking the material gradient of the placement in \eqref{eq:first_order_localization} and using the definition of $\boldsymbol{F}$ in \eqref{eq:cosserat_strain_measures} gives
\begin{align}
    \boldsymbol{F} = \bar{\boldsymbol{F}} + \operatorname{Grad} \tilde{\boldsymbol{u}} \,.
    \label{eq:gradient_localization}%
\end{align}
Within the auxiliary representation $\mathcal{C}$, consistency of the localization requires that the volume average of the microscopic deformation gradient recover the prescribed macroscopic one,
\begin{align}
    \bar{\boldsymbol{F}} = \frac{1}{V} \int_{\Omega_{0}} \boldsymbol{F} \,\mathrm{d} V \,.
    \label{eq:macro_deformation_gradient_average}%
\end{align}
Substitution of \eqref{eq:gradient_localization} into \eqref{eq:macro_deformation_gradient_average}, followed by the divergence theorem, shows that this requirement is equivalent to a vanishing mean fluctuation gradient,
\begin{align}
    \frac{1}{V} \int_{\Omega_{0}} \operatorname{Grad} \tilde{\boldsymbol{u}} \,\mathrm{d} V = \frac{1}{V} \int_{\partial \Omega_{0}} \tilde{\boldsymbol{u}} \otimes \boldsymbol{N} \,\mathrm{d} A = \boldsymbol{0} \,,
\end{align}
which can be satisfied by vanishing or periodic displacement fluctuations on the paired boundary parts $\Gamma_{i}^{\pm}$.

The microrotation field associated with the power-equivalent representation of Section~\ref{subsec:cosserat_equivalence} is localized multiplicatively using the macroscopic polar factor $\bar{\boldsymbol{R}}$,
\begin{align}
    \boldsymbol{Q}(\boldsymbol{X}) = \bar{\boldsymbol{R}} \tilde{\boldsymbol{Q}}(\boldsymbol{X}) \,,
    \label{eq:rotation_localization}%
\end{align}
where $\tilde{\boldsymbol{Q}}$ is the relative microscopic rotation. Since $\bar{\boldsymbol{R}}$ is constant over a first-order RVE, this factorization separates the uniform macroscopic rotation associated with $\bar{\boldsymbol{F}}$ from the relative microscopic rotation. Consequently, $\bar{\boldsymbol{R}}$ introduces no macroscopic curvature, and all spatial variations of $\boldsymbol{Q}$ over the RVE remain microscopic.

For the structural RVE $\mathcal{S}$, the Hill--Mandel macrohomogeneity condition \citep{hill_1963, hill_1972, mandel_1972, miehe_2003} requires equality of the macroscopic virtual stress power and the internal virtual work of the RVE per unit cell volume,
\begin{align}
    \bar{\boldsymbol{P}} : \delta \bar{\boldsymbol{F}} = \frac{1}{V} \delta W^{\mathrm{int}}_{\mathcal{S}} \,,
    \label{eq:hill_mandel_structural}%
\end{align}
where $\bar{\boldsymbol{P}}$ is the effective first Piola--Kirchhoff stress, work-conjugate to $\bar{\boldsymbol{F}}$. By the power equivalence in \eqref{eq:cosserat_power_equivalence}, the structural work equals that of the auxiliary Cosserat continuum $\mathcal{C}$. Its principle of virtual work \eqref{eq:cosserat_virtual_work_principle} and boundary-work expression \eqref{eq:cosserat_external_virtual_work} therefore transform \eqref{eq:hill_mandel_structural} into
\begin{align}
    \bar{\boldsymbol{P}} : \delta \bar{\boldsymbol{F}} = \frac{1}{V} \int_{\partial \Omega_{0}} \bigl( \boldsymbol{t} \cdot \delta \boldsymbol{u} + \boldsymbol{c} \cdot \boldsymbol{\eta} \bigr) \,\mathrm{d} A \,.
    \label{eq:hill_mandel_boundary_work}%
\end{align}
The variation of the displacement localization \eqref{eq:first_order_localization} is $\delta \boldsymbol{u} = \delta \bar{\boldsymbol{F}} \boldsymbol{X} + \delta \tilde{\boldsymbol{u}}$, and the affine contribution can be rewritten using $\boldsymbol{t} \cdot (\delta \bar{\boldsymbol{F}} \boldsymbol{X}) = (\boldsymbol{t} \otimes \boldsymbol{X}) : \delta \bar{\boldsymbol{F}}$. Substituting these two relations into \eqref{eq:hill_mandel_boundary_work} and collecting the terms conjugate to $\delta \bar{\boldsymbol{F}}$ gives the Hill--Mandel residual
\begin{align}
        0 = \left( \frac{1}{V} \int_{\partial \Omega_{0}} \boldsymbol{t} \otimes \boldsymbol{X} \,\mathrm{d} A - \bar{\boldsymbol{P}} \right) : \delta \bar{\boldsymbol{F}} + \frac{1}{V} \int_{\partial \Omega_{0}} \bigl( \boldsymbol{t} \cdot \delta \tilde{\boldsymbol{u}} + \boldsymbol{c} \cdot \boldsymbol{\eta} \bigr) \,\mathrm{d} A \,.
    \label{eq:hill_mandel_residual}%
\end{align}

A sufficient set of admissible boundary conditions follows by making the second term in \eqref{eq:hill_mandel_residual} vanish. On each pair of matching boundary points, impose the periodic kinematic constraints $\tilde{\boldsymbol{u}}^{+} = \tilde{\boldsymbol{u}}^{-}$ and $\tilde{\boldsymbol{Q}}^{+} = \tilde{\boldsymbol{Q}}^{-}$. The displacement localization \eqref{eq:first_order_localization} and the rotation factorization \eqref{eq:rotation_localization} then give, respectively,
\begin{subequations}
    \label{eq:periodic_kinematic_conditions}%
    \begin{align}
        \boldsymbol{u}^{+} - \boldsymbol{u}^{-} &= ( \bar{\boldsymbol{F}} - \boldsymbol{I} )( \boldsymbol{X}^{+} - \boldsymbol{X}^{-} ) \,,\label{eq:periodic_displacement_bc}%
        \\
        \boldsymbol{Q}^{+} - \boldsymbol{Q}^{-} &= \boldsymbol{0} \,.\label{eq:periodic_rotation_bc}%
    \end{align}
    \label{eq:periodic_bc}%
\end{subequations}
By the kinematic correspondence established in Section~\ref{subsec:cosserat_equivalence}, these conditions are imposed in the structural RVE $\mathcal{S}$ on the translational degrees of freedom of each pair of matching boundary points or nodes and, where present, on the rotational degrees of freedom of beams and shells. Admissible variations of this constraint satisfy $\delta \boldsymbol{Q}^{+} = \delta \boldsymbol{Q}^{-}$ and the virtual-spin definition \eqref{eq:virtual_spin_definition} gives $\boldsymbol{\eta}^{+} = \boldsymbol{\eta}^{-}$. The generalized force and couple resultants are not prescribed independently. Rather, stationarity of the periodically constrained problem yields the corresponding anti-periodic reaction tractions,
\begin{subequations}
    \label{eq:antiperiodic_generalized_tractions}%
    \begin{align}
        \boldsymbol{t}^{+} + \boldsymbol{t}^{-} &= \boldsymbol{0} \,,\\
        \boldsymbol{c}^{+} + \boldsymbol{c}^{-} &= \boldsymbol{0} \,.
    \end{align}
\end{subequations}

\begin{remark}[Vanishing virtual fluctuation work]
    To verify the cancellation in~\eqref{eq:hill_mandel_residual}, map each minus face onto its matching plus face by the cell translation. The map preserves the reference area measure, and the paired fluctuation work becomes
    \begin{align}
        \begin{split}
            \int_{\Gamma_{i}^{+}} \bigl( ( \boldsymbol{t}^{+} + \boldsymbol{t}^{-} ) \cdot \delta \tilde{\boldsymbol{u}}^{+} + ( \boldsymbol{c}^{+} + \boldsymbol{c}^{-} ) \cdot \boldsymbol{\eta}^{+} \bigr) \,\mathrm{d} A = 0 \,.
        \end{split}
    \end{align}
    Summing over the three face pairs therefore gives
    \begin{align}
        \frac{1}{V} \int_{\partial \Omega_{0}} \bigl( \boldsymbol{t} \cdot \delta \tilde{\boldsymbol{u}} + \boldsymbol{c} \cdot \boldsymbol{\eta} \bigr) \,\mathrm{d} A = 0 \,.
      \label{eq:zero_boundary_fluctuation_work}%
    \end{align}
    Hence the periodic constraints~\eqref{eq:periodic_kinematic_conditions}, together with their anti-periodic reaction tractions~\eqref{eq:antiperiodic_generalized_tractions}, satisfy the first-order Hill--Mandel condition within the scale-separated, first-order setting introduced above.
\end{remark}

\begin{remark}[Affine displacement and rotational Dirichlet boundary conditions]
    Besides the periodic conditions~\eqref{eq:periodic_kinematic_conditions}, fully affine displacement boundary conditions combined with rotational Dirichlet data also satisfy the first-order Hill--Mandel condition, provided that the admissible variations obey $\delta \tilde{\boldsymbol{u}} = \boldsymbol{0}$ and $\boldsymbol{\eta} = \boldsymbol{0}$ on $\partial \Omega_{0}$. Both contributions to the generalized boundary fluctuation work in \eqref{eq:hill_mandel_residual} then vanish. These boundary conditions are not discussed further, and periodic boundary conditions are employed throughout this work.
\end{remark}

With the generalized fluctuation work eliminated by \eqref{eq:zero_boundary_fluctuation_work}, the arbitrariness of $\delta \bar{\boldsymbol{F}}$ in \eqref{eq:hill_mandel_residual} requires its work-conjugate coefficient to vanish. This directly provides a representation of the effective stress $\bar{\boldsymbol{P}}$ introduced in \eqref{eq:hill_mandel_structural},
\begin{align}
    \bar{\boldsymbol{P}} = \frac{1}{V} \int_{\partial \Omega_{0}} \boldsymbol{t} \otimes \boldsymbol{X} \,\mathrm{d} A \,,
  \label{eq:mean_stress}%
\end{align}
which is the common force virial over the reference boundary and constitutes the primary definition of the effective stress, requiring only boundary data \citep{zohdi_2017}. By the kinematic identification and power equivalence of Section~\ref{subsec:cosserat_equivalence}, the auxiliary traction $\boldsymbol{t}$ is represented on the structural RVE $\mathcal{S}$ by the generalized boundary forces work-conjugate to the corresponding structural translations. Equation~\eqref{eq:mean_stress} can therefore be evaluated for RVEs comprising continuum, truss, beam, or shell constituents directly from their boundary force resultants, without reconstructing a pointwise traction or stress field on the structural constituents. For the regular auxiliary force-stress field $\boldsymbol{P}$, equilibrium $\operatorname{Div} \boldsymbol{P} = \boldsymbol{0}$ in $\Omega_{0}$ and $\boldsymbol{t} = \boldsymbol{P} \boldsymbol{N}$ allow the divergence theorem to return the same value as the equivalent volume average
\begin{align}
    \bar{\boldsymbol{P}} = \frac{1}{V} \int_{\Omega_{0}} \boldsymbol{P} \,\mathrm{d} V \,.
\end{align}
Because the macroscale body is modeled as a first-order Cauchy continuum, its power has no independent curvature term. Consequently, the couple tractions do not contribute a separate term to the first-order macroscopic stress. Nevertheless, moments remain essential to microscopic equilibrium and influence the effective response indirectly through the resulting boundary forces.

The macroscopic Cauchy stress is obtained by the push-forward
\begin{align}
    \bar{\boldsymbol{\sigma}} = \frac{1}{\bar{J}} \bar{\boldsymbol{P}} \bar{\boldsymbol{F}}^{\mathrm{T}} \,,\qquad \bar{J} \coloneqq \det \bar{\boldsymbol{F}} \,.
    \label{eq:cauchy_from_P}%
\end{align}
The effective Cauchy stress is symmetric, $\bar{\boldsymbol{\sigma}} = \bar{\boldsymbol{\sigma}}^{\mathrm{T}}$, as required for a first-order Cauchy macro-continuum. This follows from the periodic boundary conditions and the global angular-momentum balance of the equivalent Cosserat continuum. The detailed proof is given in \ref{app:symmetry}.

% ----------------------------------------------------------------------------------------
\subsection{Small-strain specialization}\label{subsec:smallstrain}%
The macroscopic displacement gradient is $\bar{\boldsymbol{H}} \coloneqq \bar{\boldsymbol{F}} - \boldsymbol{I}$, so that $\bar{\boldsymbol{F}} = \boldsymbol{I} + \bar{\boldsymbol{H}}$ holds identically. The geometrically linear regime is the specialization to a small displacement gradient,
\begin{align}
    \lVert \bar{\boldsymbol{H}} \rVert \ll 1 \,,
\end{align}
and a small auxiliary microrotation, for which $\boldsymbol{Q} \approx \boldsymbol{I} + \widehat{\boldsymbol{\theta}}$ with the infinitesimal director-rotation vector $\boldsymbol{\theta}$. The macroscopic strain is the symmetric part $\bar{\boldsymbol{\varepsilon}} \coloneqq \operatorname{sym} \bar{\boldsymbol{H}}$, where $\operatorname{sym}( \boldsymbol{A} ) \coloneqq \tfrac{1}{2} ( \boldsymbol{A} + \boldsymbol{A}^{\mathrm{T}} )$, while the complementary part $\operatorname{skew} \bar{\boldsymbol{H}}$ is an infinitesimal rigid rotation of the cell that induces neither strain nor stress. A pure macro-strain loading therefore constrains only $\bar{\boldsymbol{\varepsilon}}$ and leaves this rotation free. Choosing it to vanish makes the prescribed macroscopic displacement gradient symmetric, $\bar{\boldsymbol{H}} = \bar{\boldsymbol{\varepsilon}}$.

The reference and current configurations coincide to first order, so that \eqref{eq:cauchy_from_P} gives $\bar{\boldsymbol{P}} \approx \bar{\boldsymbol{\sigma}}$, and the reference-boundary virial~\eqref{eq:mean_stress} directly returns the Cauchy stress. The periodic conditions~\eqref{eq:periodic_kinematic_conditions} likewise linearize, the translational jump carrying the macroscopic strain and the rotational jump vanishing,
\begin{subequations}
    \begin{align}
        \boldsymbol{u}^{+} - \boldsymbol{u}^{-} &= \bar{\boldsymbol{\varepsilon}} ( \boldsymbol{X}^{+} - \boldsymbol{X}^{-} ) \,,\\
        \boldsymbol{\theta}^{+} - \boldsymbol{\theta}^{-} &= \boldsymbol{0} \,,
    \end{align}
\end{subequations}
and the deformation-gradient average~\eqref{eq:macro_deformation_gradient_average} reduces to the symmetric strain average
\begin{align}
    \bar{\boldsymbol{\varepsilon}} = \frac{1}{V} \int_{\Omega_{0}} \boldsymbol{\varepsilon} \,\mathrm{d} V = \frac{1}{V} \int_{\partial \Omega_{0}} \operatorname{sym}( \boldsymbol{u} \otimes \boldsymbol{N} ) \,\mathrm{d} A \,,
\end{align}
with $\boldsymbol{\varepsilon} \coloneqq \operatorname{sym}(\operatorname{Grad} \boldsymbol{u})$.

For linear-elastic constituents the homogenized response is
\begin{align}
  \bar{\boldsymbol{\sigma}} = \bar{\mathbb{C}} : \bar{\boldsymbol{\varepsilon}} \,,
\end{align}
with the fourth-order effective stiffness tensor $\bar{\mathbb{C}}$. Assuming a stress-free reference configuration, we identify the effective stiffness in Voigt notation by collecting the symmetric strain and stress components, respectively, as
\begin{subequations}
    \begin{align}
        \bar{\boldsymbol{\varepsilon}}^{\mathrm{V}} &\coloneqq \begin{bmatrix}
            \bar{\varepsilon}_{11} & \bar{\varepsilon}_{22} & \bar{\varepsilon}_{33} & 2 \bar{\varepsilon}_{23} & 2 \bar{\varepsilon}_{13} & 2 \bar{\varepsilon}_{12}
        \end{bmatrix}^{\mathrm{T}} \,,\\
        \bar{\boldsymbol{\sigma}}^{\mathrm{V}} &\coloneqq \begin{bmatrix}
            \bar{\sigma}_{11} & \bar{\sigma}_{22} & \bar{\sigma}_{33} & \bar{\sigma}_{23} & \bar{\sigma}_{13} & \bar{\sigma}_{12}
        \end{bmatrix}^{\mathrm{T}} \,.
    \end{align}
\end{subequations}
If an initial residual stress is present, its homogenized value must first be subtracted from each stress response. With these definitions, $\bar{\boldsymbol{\sigma}}^{\mathrm{V}} = \bar{\mathbb{C}}^{\mathrm{V}} \bar{\boldsymbol{\varepsilon}}^{\mathrm{V}}$, where $\bar{\mathbb{C}}^{\mathrm{V}} \in \mathbb{R}^{6 \times 6}$ is the Voigt stiffness matrix. Its columns are obtained one at a time by solving the boundary value problem on $\Omega_{0}$ for the six load cases
\begin{align}
    \bar{\boldsymbol{\varepsilon}}^{\mathrm{V},(\alpha)} = \varepsilon^{*} \, \boldsymbol{e}_{\alpha}^{(6)} \,,\qquad \alpha = 1, \ldots, 6 \,,
  \label{eq:independent_strain_cases_3d}%
\end{align}
with $\{\boldsymbol{e}_{\alpha}^{(6)}\}_{\alpha=1}^{6}$ the canonical basis of $\mathbb{R}^{6}$ and the amplitude $\varepsilon^{*} \ne 0$. Homogenizing the stress response yields $\bar{\boldsymbol{\sigma}}^{\mathrm{V},(\alpha)}$, and the effective stiffness follows as
\begin{align}
    \bar{\mathbb{C}}^{\mathrm{V}} = \frac{1}{\varepsilon^{*}} \begin{bmatrix}
        \bar{\boldsymbol{\sigma}}^{\mathrm{V},(1)} & \cdots & \bar{\boldsymbol{\sigma}}^{\mathrm{V},(6)}
  \end{bmatrix} \,.
  \label{eq:stiffness_columns}%
\end{align}
For materially nonlinear or history-dependent constituents no constant stiffness exists. The Hill--Mandel condition~\eqref{eq:hill_mandel_structural} and the stress virial~\eqref{eq:mean_stress} nevertheless still apply, and the homogenized response is the evolving relation between the prescribed macroscopic loading and the stress history.

% ========================================================================================
\section{Discrete formulation and implementation}\label{sec:software}%
This section casts the homogenization framework of Section~\ref{sec:homogenization} in a discrete finite element setting. Our implementation builds on the commercial finite element solver Abaqus/Standard \citep{abaqus_2025}, which supplies the element formulations for the continuum, truss, beam, and shell constituents, their constitutive models, constraint mechanisms, and the nonlinear solution procedure. The discrete scale transition requires a periodic pairing of the boundary nodes and constraint equations that impose the macroscopic deformation together with periodicity of the nodal rotations wherever rotational degrees of freedom are present. In addition, redundant periodic copies of boundary-supported elements are removed before equilibrium is solved, providing a variationally equivalent realization of the ownership-corrected internal virtual work introduced in Section~\ref{subsec:structural_rve}. The effective stress is subsequently recovered from the nodal forces of the retained elements.

Section~\ref{subsec:bcs} develops the discrete macroscopic loading and periodic constraints. Section~\ref{subsec:element_reduction} introduces the ownership-consistent periodic element reduction, and Section~\ref{subsec:stress} derives the element-type-independent effective stress recovery on the reduced model. Section~\ref{subsec:implementation} describes the computational realization.

% ----------------------------------------------------------------------------------------
\subsection{Discrete macroscopic loading and periodic constraints}\label{subsec:bcs}%
The RVE $\mathcal{S}$ is discretized by a periodic finite element mesh $\mathcal{S}_{h}$ whose nodes, indexed by $j$, carry the translations $\boldsymbol{u}_{j}$ and, at beam and shell nodes, the finite rotation vectors $\boldsymbol{\phi}_{j}$. At nodes without rotational degrees of freedom, the formal zero extensions $\boldsymbol{\phi}_{j} = \delta\boldsymbol{\phi}_{j} = \boldsymbol{\eta}_{j} = \boldsymbol{0}$ are adopted, and the corresponding nodal moment resultants are understood to vanish. Each finite rotation vector parametrizes the corresponding auxiliary microrotation $\boldsymbol{Q}$ exactly through the exponential map \citep{argyris_1982, simo_1985}
\begin{align}
    \boldsymbol{Q}(\boldsymbol{\phi}) = \exp(\widehat{\boldsymbol{\phi}}) \,.
\end{align}
For an admissible variation $\delta\boldsymbol{\phi}_{j}$, the continuous virtual-spin definition~\eqref{eq:virtual_spin_definition} gives the corresponding nodal spatial virtual rotation as
\begin{align}
    \widehat{\boldsymbol{\eta}}_{j} \coloneqq \delta\boldsymbol{Q}_{j}\boldsymbol{Q}_{j}^{\mathrm{T}} \,,\qquad \boldsymbol{\eta}_{j} = \boldsymbol{J}_{\mathrm{l}}( \boldsymbol{\phi}_{j} ) \delta\boldsymbol{\phi}_{j} \,,
    \label{eq:nodal_virtual_rotation}%
\end{align}
where $\boldsymbol{Q}_{j} \coloneqq \boldsymbol{Q}(\boldsymbol{\phi}_{j})$ and, with $\vartheta \coloneqq \lVert\boldsymbol{\phi}\rVert$, the left Jacobian of the exponential map is
\begin{align}
    \boldsymbol{J}_{\mathrm{l}} ( \boldsymbol{\phi} ) \coloneqq \boldsymbol{I} + \frac{1-\cos\vartheta}{\vartheta^{2}} \widehat{\boldsymbol{\phi}} + \frac{\vartheta-\sin\vartheta}{\vartheta^{3}} \widehat{\boldsymbol{\phi}}^{2} \,.
\end{align}
At $\boldsymbol{\phi} = \boldsymbol{0}$, the Jacobian is defined by its continuous extension $\boldsymbol{J}_{\mathrm{l}} ( \boldsymbol{0} ) = \boldsymbol{I}$. In the geometrically linear regime, Taylor expansion of the exponential map gives $\boldsymbol{Q} \approx \boldsymbol{I} + \widehat{\boldsymbol{\phi}}$, so the finite rotation vector coincides with its infinitesimal counterpart $\boldsymbol{\theta}$. 

The macroscopic loading is prescribed as a history in the pseudo-time $\tau$ that parametrizes the quasi-static response and is represented by the displacement gradient $\bar{\boldsymbol{H}} ( \tau ) = \bar{\boldsymbol{F}} ( \tau ) - \boldsymbol{I}$. As assumed in Section~\ref{subsec:smallstrain}, the skew part of $\bar{\boldsymbol{H}}$ is set to zero for geometrically linear macroscopic strain loading, so that $\bar{\boldsymbol{H}} ( \tau ) = \bar{\boldsymbol{\varepsilon}} ( \tau )$. Prescribing the history at discrete values of $\tau$ admits proportional, multiaxial, and cyclic loading paths.

\begin{figure*}[htbp]
    \centering
    \includegraphics[width=0.95\textwidth]{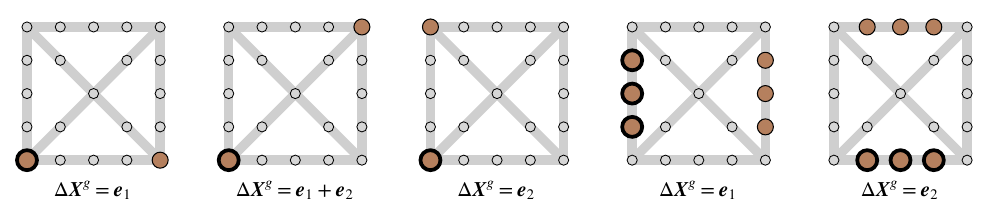}%
    \caption{Node-pair sets $\mathcal{P}_{g}$ for a square unit cell with edge length $\ell = 1$. Each panel shows one primary--secondary boundary-entity pair, with primary nodes drawn larger than secondary nodes. The first and fourth panels, as well as the third and fifth panels, share a cell translation but couple entities of different dimensionality and therefore define distinct pair sets.}
    \label{fig:periodic_groups}%
\end{figure*}

Under macroscopic loading, the periodic conditions~\eqref{eq:periodic_bc} carry a nonhomogeneous displacement jump between matched boundary points, and the loading leaves the rigid-body translation of the cell undetermined. The jump is enforced through a periodic pairing of the boundary nodes. Each boundary node is first assigned to exactly one boundary entity, namely the unit-cell face (2D), edge (1D), or corner vertex (0D) on which it lies. Edge nodes are treated as separate from the adjoining faces, and corner nodes are treated as separate from the adjoining edges. Let
\begin{align}
    \mathcal{G} \coloneqq \bigl\{ ( \mathcal{E}_{\mathrm{p}}^{g}, \mathcal{E}_{\mathrm{s}}^{g} ) \bigr\}_{g=1}^{n_{\mathrm{pairs}}} \,,
\end{align}
denote the predefined pairs of primary and secondary boundary entities. The two entities in each pair have the same dimensionality and are related by the cell translation
\begin{subequations}
    \begin{align}
        \mathcal{E}_{\mathrm{s}}^{g} &= \mathcal{E}_{\mathrm{p}}^{g} + \Delta\boldsymbol{X}^{g} \,,\\
        \Delta\boldsymbol{X}^{g} &= \sum_{i=1}^{3} c^{g}(i) \Delta\boldsymbol{X}^{i} \,,\\
        c^{g}(i) &\in \{-1,0,1\} \,,
    \end{align}
\end{subequations}
where $\Delta\boldsymbol{X}^{i}$ are the primitive cell translations. Let $\mathcal{N}_{\mathrm{p}}^{g} \coloneqq \mathcal{N}(\mathcal{E}_{\mathrm{p}}^{g})$ and $\mathcal{N}_{\mathrm{s}}^{g} \coloneqq \mathcal{N}(\mathcal{E}_{\mathrm{s}}^{g})$ denote the nodes assigned to the two boundary entities. The node-pair set associated with the entity pair $g$ is
\begin{align}
    \mathcal{P}_{g} \coloneqq \left\{ (p,s) \in \mathcal{N}_{\mathrm{p}}^{g} \times \mathcal{N}_{\mathrm{s}}^{g} \;\middle|\; \boldsymbol{X}_{s} - \boldsymbol{X}_{p} = \Delta\boldsymbol{X}^{g} \right\} \,.
    \label{eq:pair_sets}%
\end{align}
Thus, each $\mathcal{P}_{g}$ contains all matching nodes on one predefined primary--secondary entity pair. Every secondary node has exactly one primary partner, whereas a primary node may participate in several groups and thereby be tied to several secondaries. We visualize this construction in Figure~\ref{fig:periodic_groups}.

For every $(p,s) \in \mathcal{P}_{g}$, the discrete counterpart of the periodic displacement condition~\eqref{eq:periodic_displacement_bc} is
\begin{align}
    \boldsymbol{u}_{s} ( \tau ) - \boldsymbol{u}_{p} ( \tau ) = \bar{\boldsymbol{H}}( \tau ) \Delta\boldsymbol{X}^{g} \,.
    \label{eq:discrete_pair_jump}%
\end{align}
Finite element implementations commonly enforce periodic displacement jumps through linear constraint equations and auxiliary control variables \citep{tyrus_2007, omairey_2019}. Here, one auxiliary reference point is introduced for each entity pair. Its translational degrees of freedom store the macroscopic jump,
\begin{align}
    \boldsymbol{u}^{\mathrm{RP},g} ( \tau ) = \bar{\boldsymbol{H}} ( \tau ) \Delta\boldsymbol{X}^{g} \,.
    \label{eq:discrete_jump}%
\end{align}
The coupled node pairs then satisfy the homogeneous multi-point equation
\begin{align}
    \boldsymbol{u}_{s}(\tau)-\boldsymbol{u}_{p}(\tau)-\boldsymbol{u}^{\mathrm{RP},g}(\tau) = \boldsymbol{0} \,, \qquad \forall\, (p,s) \in \mathcal{P}_{g} \,.
    \label{eq:discrete_periodic_translation}%
\end{align}
Substitution of the prescribed reference-point displacement~\eqref{eq:discrete_jump} into the multi-point equation~\eqref{eq:discrete_periodic_translation} recovers the nodal jump~\eqref{eq:discrete_pair_jump}. Because each secondary node $s$ has a single primary partner $p$, nodes on edges and corner vertices are not over-constrained where periodic boundaries meet. For the cuboidal cells considered here, the node-pair translations include the three primitive cell translations and hence span the three lattice directions. A rigid rotation would alter at least one pair difference $\boldsymbol{u}_{s}-\boldsymbol{u}_{p}$ prescribed by~\eqref{eq:discrete_periodic_translation}, whereas a common rigid translation leaves all pair differences unchanged. Consequently, among the six global rigid-body modes of the three-dimensional cell, the three rotations are suppressed while the three translations remain unconstrained. Prescribing the affine displacement at a single anchor node $A$,
\begin{align}
    \boldsymbol{u}_{A} ( \tau ) = \bar{\boldsymbol{H}} ( \tau ) \boldsymbol{X}_{A} \,,
    \label{eq:discrete_anchor}%
\end{align}
removes these remaining translational modes. The anchor is chosen so that it is not the secondary node of any periodic equation and therefore does not over-constrain the system.

In contrast to the displacement, the nodal rotation field carries no macroscopic jump, so the coupled node pairs that carry rotational degrees of freedom satisfy the homogeneous multi-point equation with vanishing right-hand side,
\begin{align}
  \boldsymbol{\phi}_{s} ( \tau ) - \boldsymbol{\phi}_{p} ( \tau ) = \boldsymbol{0} \,, \qquad \forall\, (p,s) \in \mathcal{P}_{g} \,.
  \label{eq:discrete_periodic_rotation}%
\end{align}
The reference director frame is fixed and independent of position, so all nodal rotation vectors share the same global parametrization. The paired rotation vectors are assumed to remain on a common continuous branch of the exponential parametrization throughout the loading history. Equal rotation vectors in~\eqref{eq:discrete_periodic_rotation} therefore give $\boldsymbol{Q}_{s}=\boldsymbol{Q}_{p}$. Their admissible variations satisfy $\delta\boldsymbol{\phi}_{s} = \delta \boldsymbol{\phi}_{p}$, and the corresponding left Jacobians in~\eqref{eq:nodal_virtual_rotation} coincide. It follows that $\boldsymbol{\eta}_{s} = \boldsymbol{\eta}_{p}$. Hence~\eqref{eq:discrete_periodic_rotation} realizes the continuous periodic rotation condition~\eqref{eq:periodic_rotation_bc} and reduces to $\boldsymbol{\theta}_{s} = \boldsymbol{\theta}_{p}$ in the geometrically linear limit. The constraint reaction forces and moments associated with~\eqref{eq:discrete_periodic_translation} and~\eqref{eq:discrete_periodic_rotation} act with opposite signs on the primary and secondary nodes of each pair. This pairwise anti-periodicity is the discrete counterpart of the anti-periodic generalized tractions~\eqref{eq:antiperiodic_generalized_tractions}.

A boundary node whose translational degrees of freedom are already eliminated by an inter-constituent coupling, for example a beam fiber embedded in a host with a continuum element formulation, is omitted from the corresponding periodic translational equation. Provided that paired embedded nodes are coupled to periodically corresponding locations of the host discretization, the host field supplies the same macroscopic displacement jump, and their translational periodicity is therefore inherited from the host. Any rotational degrees of freedom they carry remain subject to the rotational periodic constraint.

% ----------------------------------------------------------------------------------------
\subsection{Ownership-consistent periodic element reduction}\label{subsec:element_reduction}%
The ownership correction~\eqref{eq:ownership_corrected_internal_work} must enter the discrete boundary-value problem. For an element $e \in \mathcal{S}_{h}$ with nodes $j \in e$, define
\begin{align}
    \mathcal{F}_{e} \coloneqq \left\{ \Gamma_{i}^{\varsigma} \;\middle|\; i \in \{1,2,3\},\; \varsigma \in \{+,-\},\; \boldsymbol{X}_{j} \in \Gamma_{i}^{\varsigma}\; \forall\, j \in e \right\} \,.
\end{align}
For the boundary-conforming isoparametric mesh, this nodal criterion identifies $\mathcal{F}_{e}$ as the elementwise set of cell faces containing the complete material support of $e$, in accordance with the pointwise definition~\eqref{eq:boundary_face_set}. Its cardinality determines the discrete ownership factor
\begin{align}
    \omega_{e} \coloneqq 2^{-\lvert\mathcal{F}_{e}\rvert} \,.
\end{align}
Thus, fractional ownership applies only to elements whose complete material support lies on one or more cell faces. An element that merely terminates at or crosses the cell boundary satisfies $\mathcal{F}_{e} = \varnothing$ and therefore retains unit ownership. For each element with $\mathcal{F}_{e} \neq \varnothing$, the primary--secondary boundary-entity pairs collected in $\mathcal{G}$ identify its periodically translated copies in the complete mesh. Since the discretization, constitutive properties, and initial material state are periodic, the constraints ensure that these copies undergo identical deformation and rotation histories and hence have identical internal virtual-work histories. Each such periodic copy set comprises $\omega_{e}^{-1}$ elements, so its ownership-weighted total equals the unweighted contribution of any one copy.

\begin{figure*}[htbp]
    \centering
    \includegraphics[width=0.65\textwidth]{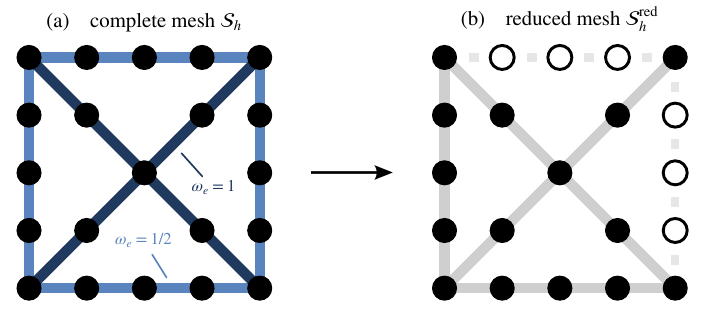}
    \caption{Ownership-consistent periodic element reduction for a two-dimensional unit cell discretized with truss or beam elements: (a) complete periodic mesh $\mathcal{S}_{h}$; (b) reduced mesh $\mathcal{S}_{h}^{\mathrm{red}}$, in which the copies assigned to secondary boundary entities are removed while all nodes are retained.}
    \label{fig:element_reduction}%
\end{figure*}

This observation permits the ownership correction to be realized directly by element reduction. For each periodic copy set, the implementation retains one representative according to the predefined primary--secondary assignment and removes the copies assigned to secondary boundary entities, as illustrated in Figure~\ref{fig:element_reduction}. Let $\mathcal{S}_{h}^{\mathrm{sec}}$ collect the elements whose complete material support is assigned to a secondary boundary entity. The reduced structural model is
\begin{align}
    \mathcal{S}_{h}^{\mathrm{red}} \coloneqq \mathcal{S}_{h} \setminus \mathcal{S}_{h}^{\mathrm{sec}} \,.
    \label{eq:reduced_element_set}%
\end{align}
The reduction removes only elements. All nodes and the periodic constraint equations associated with the unreduced structural mesh are retained. Thus, the translational and rotational pairing remains unchanged. A rotation at a node orphaned by removal of its structural element copy is retained as a dependent bookkeeping degree of freedom. Since no retained element acts on this rotation, it carries neither a nodal moment nor internal virtual work. Let $\boldsymbol{f}_{j}^{(e)}$ and $\boldsymbol{m}_{j}^{(e)}$ denote the force and moment, respectively, exerted by element $e$ on its node $j$ and expressed in the global basis. With this force-on-node convention, the element internal virtual work is
\begin{align}
    \delta W_{e}^{\mathrm{int}} \coloneqq -\sum_{j \in e} \bigl( \boldsymbol{f}_{j}^{(e)} \cdot \delta\boldsymbol{u}_{j} + \boldsymbol{m}_{j}^{(e)} \cdot \boldsymbol{\eta}_{j} \bigr) \,.
    \label{eq:element_internal_virtual_work}%
\end{align}
For every admissible periodic virtual motion, the unweighted virtual work of the reduced model equals the ownership-weighted virtual work of the complete periodic mesh,
\begin{align}
    \sum_{e \in \mathcal{S}_{h}} \omega_{e} \delta W_{e}^{\mathrm{int}} = \sum_{e \in \mathcal{S}_{h}^{\mathrm{red}}} \delta W_{e}^{\mathrm{int}} \,.
\end{align}
The reduction is therefore variationally equivalent to the ownership correction. Because the equivalence holds for the complete element virtual work, it also carries over to the constrained residual and consistent tangent. The virtual work is evaluated with the total kinematic variations, while inelasticity enters through the stresses, resultants, and internal variables supplied by the history-dependent constitutive update. The identical constitutive histories of periodic copies make the reduction applicable to geometrically nonlinear, materially nonlinear, and path-dependent constituents without scaling their material parameters or internal variables.

At each pseudo-time $\tau$, the element residuals of the retained continuum, truss, beam, and shell elements are assembled, and the dependent degrees of freedom are eliminated through the periodic constraints~\eqref{eq:discrete_periodic_translation} and~\eqref{eq:discrete_periodic_rotation} and the anchor condition~\eqref{eq:discrete_anchor}. This gives the reduced quasi-static equilibrium problem
\begin{align}
    \boldsymbol{R}(\boldsymbol{q}) \coloneqq \mathop{\vcenter{\hbox{\scalebox{1.4}{$\mathcal{A}$}}}}\limits_{e \in \mathcal{S}_{h}^{\mathrm{red}}} \boldsymbol{R}_{e}(\boldsymbol{q}) = \boldsymbol{0} \,,
    \label{eq:reduced_equilibrium}%
\end{align}
where $\boldsymbol{R}_{e}$ denotes the element residual, $\mathcal{A}$ denotes the constrained finite element assembly operation, and $\boldsymbol{q}$ collects the independent nodal unknowns. The periodic equations express the secondary nodal degrees of freedom through the corresponding primary degrees of freedom and, for translations, the prescribed reference-point degrees of freedom. The finite element solver applies its standard Newton iteration to the resulting system. The reduced element topology and the constraint equations remain fixed throughout a load history, i.e., only the prescribed macroscopic data change between increments or load cases.

% ----------------------------------------------------------------------------------------
\subsection{Element-type-independent stress recovery}\label{subsec:stress}%
The effective stress is recovered directly from the converged translational element nodal forces $\boldsymbol{f}_{j}^{(e)}$. Using the paired-node sets defined above, the set of physical boundary nodes is
\begin{align}
    \mathcal{N}_{\partial} \coloneqq \bigcup_{g=1}^{n_{\mathrm{pairs}}} ( \mathcal{N}_{\mathrm{p}}^{g} \cup \mathcal{N}_{\mathrm{s}}^{g} ) \,.
\end{align}
For each $j \in \mathcal{N}_{\partial}$, let $\boldsymbol{f}_{j}^{\mathrm{reac}}$ denote the resultant translational reaction force associated with the periodic translation equations, acting on the structural RVE at node $j$. The anchor fixes only the arbitrary rigid translation and carries no reaction in an equilibrated cell, apart from the numerical solution tolerance. The periodic nodal reactions are the discrete boundary-force resultants corresponding to the traction $\boldsymbol{t}$ in \eqref{eq:mean_stress}. The discrete counterpart of the boundary virial is therefore
\begin{align}
    \bar{\boldsymbol{P}} = \frac{1}{V} \sum_{j \in \mathcal{N}_{\partial}} \boldsymbol{f}_{j}^{\mathrm{reac}} \otimes \boldsymbol{X}_{j} \,.
    \label{eq:discrete_boundary_stress}%
\end{align}
Let $\boldsymbol{f}_{j}^{\mathrm{cpl}}$ denote the resultant translational reaction from internal inter-constituent couplings at node $j$, with $\boldsymbol{f}_{j}^{\mathrm{cpl}} = \boldsymbol{0}$ at uncoupled nodes. Likewise, set $\boldsymbol{f}_{j}^{\mathrm{reac}} = \boldsymbol{0}$ for $j\notin\mathcal{N}_{\partial}$. Nodal equilibrium at every physical node can then be written as
\begin{align}
    \sum_{\substack{e \in \mathcal{S}_{h}^{\mathrm{red}}\\ e \ni j}} \boldsymbol{f}_{j}^{(e)} + \boldsymbol{f}_{j}^{\mathrm{cpl}} + \boldsymbol{f}_{j}^{\mathrm{reac}} = \boldsymbol{0} \,.
    \label{eq:nodal_equilibrium}%
\end{align}
Taking the outer product of the nodal equilibrium equation~\eqref{eq:nodal_equilibrium} with the nodal positions, summing over all physical nodes, and reordering the nodal contributions by element give
\begin{align}
    \sum_{j \in \mathcal{N}_{\partial}} \boldsymbol{f}_{j}^{\mathrm{reac}} \otimes \boldsymbol{X}_{j} = -\sum_{e \in \mathcal{S}_{h}^{\mathrm{red}}} \sum_{j \in e} \boldsymbol{f}_{j}^{(e)} \otimes \boldsymbol{X}_{j} -\sum_{j} \boldsymbol{f}_{j}^{\mathrm{cpl}} \otimes \boldsymbol{X}_{j} \,.
    \label{eq:nodal_first_moment}%
\end{align}
The internal couplings are power-consistent and reproduce affine translations. Their reactions therefore perform no virtual work under $\delta\boldsymbol{u}_{j} = \delta \bar{\boldsymbol{F}} \boldsymbol{X}_{j}$ and satisfy $\sum_{j}\boldsymbol{f}_{j}^{\mathrm{cpl}} \otimes \boldsymbol{X}_{j} = \boldsymbol{0}$. The coupling-force term in~\eqref{eq:nodal_first_moment} thus vanishes. Substitution of the remaining reaction-force virial into~\eqref{eq:discrete_boundary_stress} then gives
\begin{align}
    \bar{\boldsymbol{P}} = -\frac{1}{V} \sum_{e \in \mathcal{S}_{h}^{\mathrm{red}}} \sum_{j \in e} \boldsymbol{f}_{j}^{(e)} \otimes \boldsymbol{X}_{j} \,.
    \label{eq:discrete_stress}%
\end{align}
Nodes retained solely for the periodic constraints carry no element force from a removed element and consequently add no contribution to the effective stress. The element nodal forces are work-conjugate resultants rather than pointwise stress samples. The element virial~\eqref{eq:discrete_stress} is the finite element evaluation of the continuous effective-stress expression~\eqref{eq:mean_stress} \citep{miehe_2002, miehe_2003}. No reconstruction of a common pointwise stress field is required. The same expression applies to continuum, truss, beam, shell, and mixed RVEs because all formulation and constitutive dependence is contained in the element forces. Only translational nodal forces enter directly. Nodal moments do not appear because the first-order Cauchy macro-continuum has no independent curvature measure. They nevertheless remain essential to the microscopic equilibrium of beams and shells and influence $\bar{\boldsymbol{P}}$ through the resulting nodal forces.

At a converged constrained state, the recovered stress satisfies the discrete counterpart of the continuous Hill--Mandel condition~\eqref{eq:hill_mandel_structural},
\begin{align}
    \bar{\boldsymbol{P}} : \delta \bar{\boldsymbol{F}} = \frac{1}{V} \sum_{e \in \mathcal{S}_{h}^{\mathrm{red}}} \delta W_{e}^{\mathrm{int}} \,.
    \label{eq:discrete_hillmandel}%
\end{align}
Through the element virtual work~\eqref{eq:element_internal_virtual_work}, the microscopic side of~\eqref{eq:discrete_hillmandel} includes the moment--rotation contributions of beam and shell elements. For an a~posteriori evaluation, the finite increments between two successive converged states are inserted as a particular test direction in~\eqref{eq:discrete_hillmandel}. The spatial rotation direction associated with the finite rotation-vector increment is $\boldsymbol{\eta}_{j}^{\Delta} = \boldsymbol{J}_{\mathrm{l}}(\boldsymbol{\phi}_{j}) \Delta \boldsymbol{\phi}_{j}$, evaluated at the current state. Let $\Delta W_{e}^{\mathrm{int}}$ denote the element work~\eqref{eq:element_internal_virtual_work} evaluated with the current-state element forces and moments and the finite increments $\Delta \boldsymbol{u}_{j}$ and $\boldsymbol{\eta}_{j}^{\Delta}$. The corresponding macroscopic and microscopic incremental-work densities are
\begin{subequations}
    \begin{align}
        \Delta w_{\mathrm{macro}} &\coloneqq \bar{\boldsymbol{P}} : \Delta\bar{\boldsymbol{F}} \,,\\
        \Delta w_{\mathrm{micro}} &\coloneqq \frac{1}{V}\sum_{e \in \mathcal{S}_{h}^{\mathrm{red}}}
        \Delta W_{e}^{\mathrm{int}} \,.
    \end{align}
\end{subequations}
A normalized discrete Hill--Mandel residual is then
\begin{align}
    \epsilon_{\mathrm{HM}} \coloneqq \frac{\lvert \Delta w_{\mathrm{macro}}-\Delta w_{\mathrm{micro}}\rvert}{\max \bigl( \lvert \Delta w_{\mathrm{macro}}\rvert, \lvert \Delta w_{\mathrm{micro}}\rvert, \Delta w_{\mathrm{tol}} \bigr)} \,,
    \label{eq:hill_mandel_residual_measure}%
\end{align}
where $\Delta w_{\mathrm{tol}}>0$ is a small regularizing incremental-work-density scale. A small $\epsilon_{\mathrm{HM}}$ provides an a~posteriori check of the discrete Hill--Mandel power balance along the realized increment, including the rotational contributions.

\begin{remark}[Consistent macroscopic tangent]
The present framework is used as a stand-alone RVE analysis that returns the effective stress response for prescribed macroscopic loading paths, and no consistent macroscopic tangent $\partial\bar{\boldsymbol{P}} / \partial\bar{\boldsymbol{F}}$ is formed. Such a tangent may be required to drive the macroscopic Newton iteration of a nested FE\textsuperscript{2} computation \citep{smit_1998}. It can be obtained from the same constrained reduced model either by perturbing $\bar{\boldsymbol{F}}$ and reusing the stress recovery~\eqref{eq:discrete_stress} or by consistently linearizing the reduced equilibrium problem~\eqref{eq:reduced_equilibrium} and statically condensing it onto the auxiliary reference-point degrees of freedom \citep{kouznetsova_2001, miehe_2003}. No additional ownership correction is required because the residual and tangent of the reduced model are already variationally equivalent to those of the ownership-weighted complete cell.
\end{remark}

% ----------------------------------------------------------------------------------------
\subsection{Computational realization}\label{subsec:implementation}%
The discrete procedure is implemented in the Python library \texttt{abqhom}, which couples Gmsh \citep{geuzaine_2009} for geometry and meshing to Abaqus/Standard for the finite element solution. The input is either an axis-aligned periodic geometric model or an externally generated periodic mesh. Physical groups identify the material supports and transfer the constituent partition to the finite element model, where independent element formulations, materials, and sections can be assigned to the continuum, truss, beam, and shell constituents.

For a geometric model, the bounding box is used to classify the cell faces, edges, and vertices and to impose the translational relations required for periodic mesh generation. For an existing mesh, the boundary nodes are classified directly from their coordinates and periodicity is verified by comparing the coordinates along the free directions of each primary--secondary entity pair. In either case, every boundary node is assigned to one boundary entity and the matching coordinates determine the node-pair sets of~\eqref{eq:pair_sets}. An element is classified as boundary-supported only when all its nodes lie on the same boundary entity. The elements assigned to secondary entities are collected and removed upon transfer to the finite element model, while their nodes are retained. This realizes the reduced model~\eqref{eq:reduced_element_set} without altering the periodic nodal topology.

The macroscopic loading history is applied through the auxiliary reference points, periodic equations, and anchor condition described in Section~\ref{subsec:bcs}. If an internal coupling has already eliminated the translational degrees of freedom of a boundary node, that node is omitted from the corresponding translational periodic equation and inherits its periodicity from the host discretization. Abaqus/Standard then solves the reduced constrained problem using the selected element and constitutive formulations. At every requested converged state, the element-labeled nodal forces are retained separately and inserted into the force virial~\eqref{eq:discrete_stress}. The same post-processing operation therefore applies to the different constituent types and to linear, nonlinear, and history-dependent analyses. By prescribing general macroscopic loading paths, the framework can directly generate homogenized stress--deformation histories for calibrating conventional or neural-network-based constitutive models \citep{gorji_2020, klein_2022}. Finite element fields can additionally be exported to the VTK format using Paraqus \citep{furlan_2025}. The implementation and the scripts reproducing the numerical examples are openly available (see the data-availability statement).

% ========================================================================================
\section{Numerical examples}\label{sec:results}%
Three numerical examples assess complementary aspects of the framework: small-strain verification for a mixed continuum--shell--beam cell (Section~\ref{subsec:mixed_dimensional_verification}), finite-strain buckling of an imperfect hyperelastic Octet lattice (Section~\ref{subsec:finite_strain_octet}), and history-dependent plasticity of an elastoplastic sandwich cell (Section~\ref{subsec:elastoplastic_sandwich}).

\subsection{Small-strain verification}\label{subsec:mixed_dimensional_verification}%
The first example verifies the effective stiffness, periodic element reduction, and discrete power balance for a cell that simultaneously contains continuum, shell, and beam constituents. The cubic cell is occupied by a three-dimensional continuum. Shell midsurfaces coincide with its six faces, and beam centerlines coincide with its twelve edges. The three discretizations are conforming and share their nodes. The resulting model combines translational and rotational degrees of freedom and admits an affine closed-form solution for the normal-strain load cases. The shear load cases additionally activate the rotational response of the structural constituents.

\begin{figure}[htbp]
    \centering
    \includegraphics[width=0.95\columnwidth]{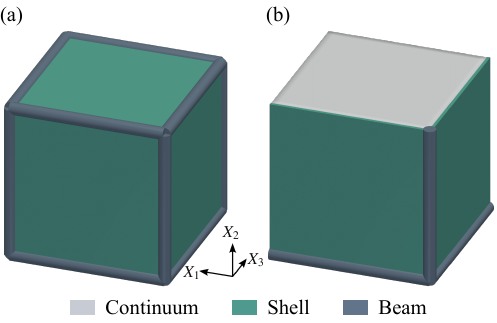}
    \caption{Mixed-dimensional verification cell: (a) complete periodic representation; (b) ownership-consistent reduced model. Colors distinguish the continuum, shell, and beam constituents.}
    \label{fig:mixed_dimensional_cell}%
\end{figure}

All geometric and material parameters are nondimensional, and the cube edge length is $\ell=1$. Figure~\ref{fig:mixed_dimensional_cell} compares the complete periodic representation with the ownership-consistent reduced model used in the analysis. The continuum mesh uses $25$ elements along each cell edge. This resolution is inherited by the shell meshes on the faces and the beam meshes along the edges. The reduction retains the continuum mesh, one shell mesh from each pair of opposite faces, and one representative from each periodic set of four beams. All nodes of the complete periodic mesh are retained so that the periodic constraints remain unchanged. The continuum, beams, and shells are discretized by eight-node trilinear Lagrange brick elements (C3D8), two-node linear Timoshenko beam elements (B31), and four-node, fully integrated, general-purpose S4 shell elements with finite membrane strains and first-order transverse-shear flexibility, respectively. The S4 formulation approaches the discrete Kirchhoff limit as the shell becomes thin \citep{abaqus_2025}. The constituents are isotropic and linearly elastic. Table~\ref{tab:example_1_parameters} summarizes their parameters and discretizations. The shell thickness is denoted by $t$, while $r$ denotes the radius of the circular beam cross-section, with area $A_{\mathrm{b}}=\pi r^{2}$.

\begin{table}[htbp]
    \centering
    \caption{Nondimensional constituent parameters and discretization. The Young's modulus is denoted $E$ and the Poisson's ratio $\nu$.}
    \label{tab:example_1_parameters}%
    \small
    \setlength{\tabcolsep}{7pt}
    \renewcommand{\arraystretch}{1.1}
    \begin{tabular}{lcccc}
        \hline
        Constituent & $E$ & $\nu$ & Section & Element \\
        \hline
        Continuum & $100$  & $0.3$ & --       & C3D8 \\
        Shell     & $500$  & $0.3$ & $t=0.02$ & S4 \\
        Beam      & $2000$ & $0.3$ & $r=0.04$ & B31 \\
        \hline
    \end{tabular}
\end{table}

The cell geometry and constituent arrangement possess cubic symmetry. The normal block of the effective stiffness matrix $\bar{\mathbb{C}}^{\mathrm{V}}$, relating the normal stresses to the normal strains, is therefore characterized by $\bar{C}_{11}^{\mathrm{V}} = \bar{C}_{22}^{\mathrm{V}}  = \bar{C}_{33}^{\mathrm{V}}$ and $\bar{C}_{12}^{\mathrm{V}}  = \bar{C}_{13}^{\mathrm{V}} = \bar{C}_{23}^{\mathrm{V}}$, with the remaining off-diagonal entries following from major symmetry. The normal-strain response provides a closed-form reference for these two independent coefficients. Let $\mathrm{c}$, $\mathrm{s}$, and $\mathrm{b}$ identify continuum, shell, and beam quantities, respectively, and let $\lambda_{\mathrm{c}}$ and $\mu_{\mathrm{c}}$ denote the Lam\'e constants of the continuum. With the continuum parameters in Table~\ref{tab:example_1_parameters}, the standard isotropic relations give $\lambda_{\mathrm{c}} = 57.692$ and $\mu_{\mathrm{c}} = 38.462$. Summing the three-dimensional continuum, plane-stress membrane, and axial beam responses with ownership weighting gives
\begin{subequations}
    \begin{align}
        \bar{C}_{11}^{\mathrm{V},\mathrm{ana}} &= \lambda_{\mathrm{c}}+2\mu_{\mathrm{c}} +\frac{2t}{\ell} \frac{E_{\mathrm{s}}}{1-\nu_{\mathrm{s}}^{2}} + \frac{A_{\mathrm{b}}}{\ell^{2}} E_{\mathrm{b}} = 166.647 \, ,\\
        \bar{C}_{12}^{\mathrm{V},\mathrm{ana}} &= \lambda_{\mathrm{c}} + \frac{t}{\ell}\frac{\nu_{\mathrm{s}} E_{\mathrm{s}}}{1-\nu_{\mathrm{s}}^{2}} = 60.989 \, .
    \end{align}
    \label{eq:example_1_analytical_stiffness}%
\end{subequations}

\begin{figure}[htbp]
    \centering
    \includegraphics[width=0.95\columnwidth]{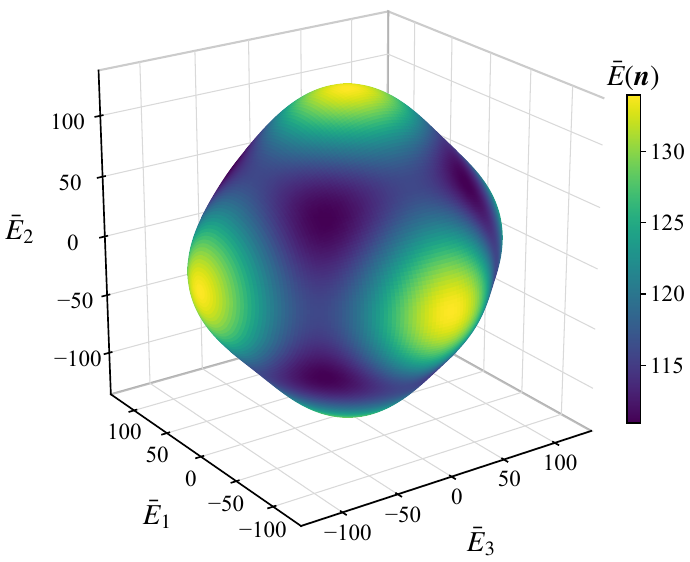}
    \caption{Directional Young's-modulus surface of the reduced mixed-dimensional cell. The radial distance and color represent $\bar{E} ( \boldsymbol{n} )$, and the Cartesian surface coordinates satisfy $\bar{E}_{i} = \bar{E} ( \boldsymbol{n} ) n_{i}$. Equal axial maxima and diagonal minima reflect the cubic symmetry of the effective stiffness.}
    \label{fig:example_1_stiffness}%
\end{figure}

Six independent macroscopic strain states are prescribed according to~\eqref{eq:independent_strain_cases_3d} with $\varepsilon^{*} = 0.01$, and the columns of the effective stiffness matrix $\bar{\mathbb{C}}^{\mathrm{V}}$ are recovered from~\eqref{eq:stiffness_columns}. Its normal block agrees with the analytical block defined by~\eqref{eq:example_1_analytical_stiffness} to a relative Frobenius-norm error of $5.3 \times 10^{-9}$. To verify the expected symmetry of the complete numerical stiffness matrix, let $\bar{\mathbb{C}}^{\mathrm{V},\mathrm{cub}}$ denote its cubic projection, obtained by averaging symmetry-equivalent entries and setting the nominally vanishing coupling entries to zero. The relative Frobenius distance is
\begin{align}
    \epsilon_{\mathrm{cub}} \coloneqq \frac{\left\lVert \bar{\mathbb{C}}^{\mathrm{V}} - \bar{\mathbb{C}}^{\mathrm{V},\mathrm{cub}} \right\rVert_{\mathrm{F}}}{\left\lVert \bar{\mathbb{C}}^{\mathrm{V}} \right\rVert_{\mathrm{F}}} = 3.8 \times 10^{-15} \,.
\end{align}
The numerical stiffness $\bar{\mathbb{C}}^{\mathrm{V}}$ is therefore cubic to numerical precision, with the three independent coefficients $\bar{C}_{11}^{\mathrm{V}} = 166.647$, $\bar{C}_{12}^{\mathrm{V}} = 60.989$, and $\bar{C}_{44}^{\mathrm{V}} = 42.421$. The corresponding directional Young's-modulus surface in Figure~\ref{fig:example_1_stiffness} provides a geometric visualization of this cubic response. Its radial distance in each direction $\boldsymbol{n}$ equals the effective modulus $\bar{E} ( \boldsymbol{n} )$. The equal maxima along the cube axes ($\bar{E} = 134.0$) and equal minima along the body diagonals ($\bar{E} = 111.0$) reflect the symmetry-equivalent directions.

The periodic element reduction is verified independently by also analyzing the complete cell with all periodic element copies retained and the Young's moduli of the shell and beam constituents scaled by their ownership factors $1/2$ and $1/4$, respectively. In the present linear-elastic setting, this directly realizes the ownership-weighted virtual work. Let $\bar{\mathbb{C}}^{\mathrm{V},\mathrm{w}}$ denote the resulting effective stiffness matrix. Its relative difference from the effective stiffness of the reduced cell is
\begin{align}
    \epsilon_{\mathrm{red}} \coloneqq \frac{\left\lVert \bar{\mathbb{C}}^{\mathrm{V}} - \bar{\mathbb{C}}^{\mathrm{V},\mathrm{w}} \right\rVert_{\mathrm{F}}}{\left\lVert \bar{\mathbb{C}}^{\mathrm{V}} \right\rVert_{\mathrm{F}}} = 7.8 \times 10^{-15} \,.
\end{align}
Thus, the two representations are numerically equivalent to machine precision. The a~posteriori power-balance check gives a maximum Hill--Mandel residual~\eqref{eq:hill_mandel_residual_measure} of $2.2 \times 10^{-8}$ over the six load cases. In particular, the shear cases provide this check while the beam and shell rotational degrees of freedom are active. The structural nodal rotations under shear, however, remain small, with a maximum magnitude $\lVert\boldsymbol{\theta}\rVert$ of $6.3\times10^{-3}$.

\subsection{Hyperelastic Octet lattice with imperfections}
\label{subsec:finite_strain_octet}%
The second example demonstrates the finite-strain framework for a purely structural RVE undergoing large rotations and a buckling-induced change in deformation mechanism. The RVE is a conventional cubic Octet lattice with edge length $\ell=1$. Its joints occupy the eight cell vertices and six face centers, and the straight struts form a compatible periodic network. All geometric and material quantities are nondimensional. The struts have the circular cross section of radius $r = 0.04$ and are discretized by 16 two-node shear-deformable beam elements (B31) along each member. The base material follows a compressible Neo--Hookean model with Young's modulus $E=100$ and Poisson's ratio $\nu=0.3$. The strain-energy density per unit reference volume is
\begin{align}
    \Psi = \frac{G}{2} \bigl( \bar{I}_{1}-3 \bigr) + \frac{K}{2} ( J-1 )^{2} \, ,
\end{align}
where
\begin{align}
    J \coloneqq \det\boldsymbol{F} \,,\qquad \boldsymbol{C} \coloneqq \boldsymbol{F}^{\mathrm{T}} \boldsymbol{F} \, ,\qquad \bar{I}_{1} \coloneqq J^{-2/3} \operatorname{tr} \boldsymbol{C} \, .
\end{align}
Here, $G$ and $K$ denote the shear and bulk moduli, respectively. Their values, $G = 38.462$ and $K = 83.333$, follow through the standard relations between isotropic elastic constants.

\begin{figure}[htbp]
    \centering
    \includegraphics[width=0.95\columnwidth]{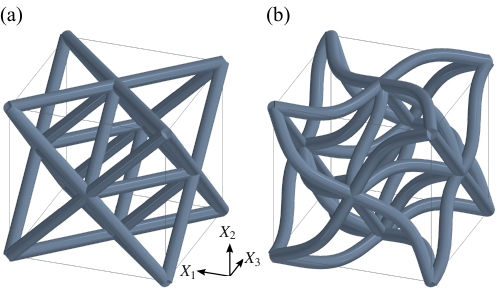}%
    \caption{Octet lattice used for the finite-strain example: (a) perfect periodic cell; (b) selected eigenvector from the lowest two-dimensional periodic buckling eigenspace, displayed with an exaggerated amplitude. The actual geometric imperfection has $a/r=0.01$.}
    \label{fig:octet_cell}%
\end{figure}

The imperfection is derived from a linearized buckling analysis of the same ownership-consistent reduced cell used in the nonlinear analyses, with all periodic nodal constraints retained. The buckling calculation uses a linear-elastic material with $E$ and $\nu$ equal to the infinitesimal elastic constants of the Neo--Hookean base material. With $-\bar{H}_{11} = 1$ as the reference load, the two lowest modes form a degenerate eigenspace with load multiplier $0.02709$. One of the two numerically returned eigenvectors is retained for the perturbation, thereby selecting a definite buckling branch. Only its translational components perturb the reference nodal coordinates, thereby defining a stress-free geometric imperfection and the corresponding initial beam tangents. The normalized translational eigenmode $\hat{\boldsymbol{u}}^{\mathrm{eig}}$ is mapped directly onto the mesh according to
\begin{align}
    \boldsymbol{X}_{j} \leftarrow \boldsymbol{X}_{j} + a \hat{\boldsymbol{u}}_{j}^{\mathrm{eig}} \,,\qquad \max_{j}\left\lVert \hat{\boldsymbol{u}}_{j}^{\mathrm{eig}} \right\rVert = 1 \, ,
\end{align}
with $a/r = 0.01$. Because the eigenproblem contains the same periodic constraints as the nonlinear RVE problem, the perturbed mesh remains periodically compatible. An independent $2 \times 2 \times 2$ supercell analysis recovers the same two-dimensional primitive-periodic lowest eigenspace, and no lower non-primitive mode is detected within this supercell. Figure~\ref{fig:octet_cell} shows the perfect cell and the selected eigenvector of the lowest periodic buckling eigenspace used to construct its imperfection.

\begin{figure*}[t]
    \centering
    \includegraphics[width=0.95\textwidth]{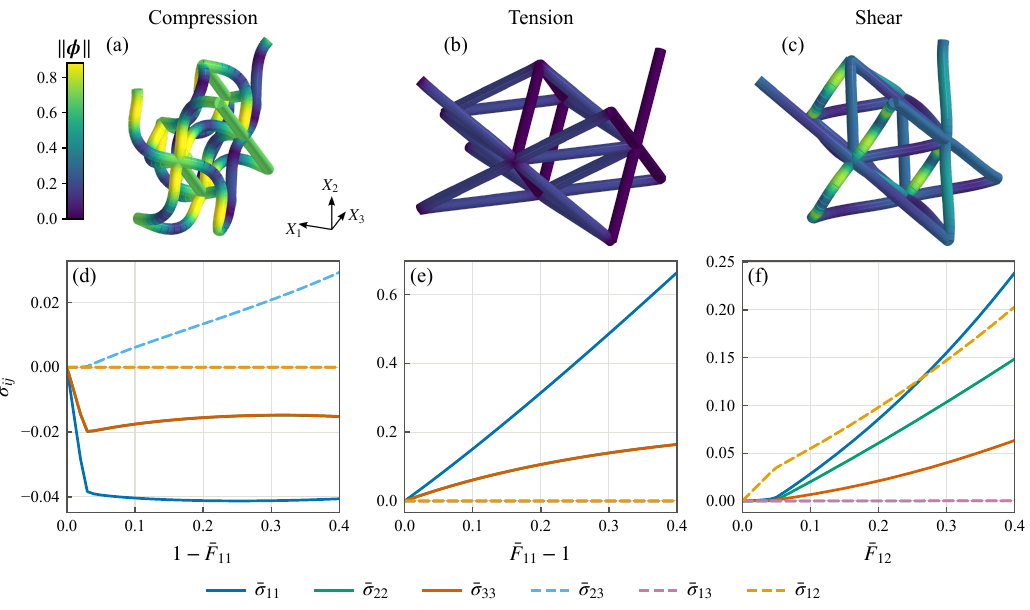}%
    \caption{Response of the imperfect Octet lattice under compression, tension, and shear. Top: the final deformed configurations colored by the nodal rotation-vector magnitude $\lVert\boldsymbol{\phi}\rVert$ for (a) compression, (b) tension, and (c) shear. Bottom: the six independent components of the macroscopic Cauchy stress against the macroscopic engineering strain for (d) compression, (e) tension, and (f) shear.}
    \label{fig:octet_response}%
\end{figure*}

Starting from the same imperfect geometry, the macroscopic deformation gradients for compression, tension, and shear, denoted by $\bar{\boldsymbol{F}}_{\mathrm{c}}$, $\bar{\boldsymbol{F}}_{\mathrm{t}}$, and $\bar{\boldsymbol{F}}_{\mathrm{s}}$, respectively, are prescribed along three proportional loading paths:
\begin{subequations}
\begin{align}
    \bar{\boldsymbol{F}}_{\mathrm{c}}(\tau) &\coloneqq \boldsymbol{I}-0.4\tau\,\boldsymbol{e}_{1}\otimes\boldsymbol{e}_{1} \,,
    \\
    \bar{\boldsymbol{F}}_{\mathrm{t}}(\tau) &\coloneqq \boldsymbol{I}+0.4\tau\,\boldsymbol{e}_{1}\otimes\boldsymbol{e}_{1} \,,
    \\
    \bar{\boldsymbol{F}}_{\mathrm{s}}(\tau) &\coloneqq \boldsymbol{I}+0.4\tau\,\boldsymbol{e}_{1}\otimes\boldsymbol{e}_{2} \,,
\end{align}
\end{subequations}
where $\tau\in[0,1]$. The paths impose 40\,\% engineering compression and tension in the $X_{1}$-direction and an engineering shear of $0.4$ in the $X_{1}$--$X_{2}$ plane. Because the normal paths constrain the transverse macroscopic strains, they represent prescribed-strain rather than uniaxial-stress tests. The boundary value problems are solved quasi-statically by implicit dynamics using the nondimensional density $\rho=2.5\times10^{-7}$. For $\tau\geq0.02$, the maximum kinetic-to-internal-energy ratio over all three analyses is $7.5\times10^{-6}$ and the maximum Hill--Mandel residual~\eqref{eq:hill_mandel_residual_measure} is $6.3\times10^{-5}$, confirming negligible inertia and an accurate discrete power balance. Beam-to-beam contact is not modeled for simplicity. The large-deformation response therefore does not capture post-contact densification of the unit cell.

Figure~\ref{fig:octet_response} shows the deformed configurations and all six independent components of the effective Cauchy stress. Under compression, the initially stretching-dominated response loses axial tangent stiffness near $1-\bar{F}_{11}=0.03$, consistent with the eigenvalue $0.02709$, as axial strut deformation gives way to coordinated bending. This geometric transition is not constitutive softening of the hyperelastic base material. The compressive stress $-\bar{\sigma}_{11}$ reaches $0.0412$ at approximately 25\,\% compression and decreases mildly to $0.0406$ at 40\,\%. Constrained transverse strains generate normal reactions, and the symmetry-breaking eigenvector produces $\bar{\sigma}_{23}$ on the post-buckling branch while $\bar{\sigma}_{12}$ and $\bar{\sigma}_{13}$ remain negligible. When the selected primitive-cell imperfection is periodically repeated over the $2 \times 2 \times 2$ supercell, the maximum relative Frobenius-norm difference between the two effective Cauchy stress tensors over the load history is $0.0026\,\%$. Under tension, $\bar{\sigma}_{11}$ increases monotonically to $0.664$, the transverse normal stresses remain nearly equal, and the shear reactions are negligible. The shear response is likewise monotonic, reaching $\bar{\sigma}_{12}=0.203$ at $\bar{F}_{12}=0.4$. Finite-strain coupling under constrained transverse deformation produces accompanying normal stresses, while $\bar{\sigma}_{13}$ and $\bar{\sigma}_{23}$ remain small. The configurations show coordinated bending under compression, predominantly axial stretching under tension, and combined axial deformation and member rotation under shear. The corresponding maximum nodal rotation-vector magnitudes $\max_{j}\lVert\boldsymbol{\phi}_{j}\rVert$ are $0.880$, $0.180$, and $0.751$, confirming that all three cases exercise finite-rotation kinematics.

\subsection{Elastoplastic sandwich cell}
\label{subsec:elastoplastic_sandwich}%
\begin{figure}[htbp]
    \centering
    \includegraphics[width=0.95\columnwidth]{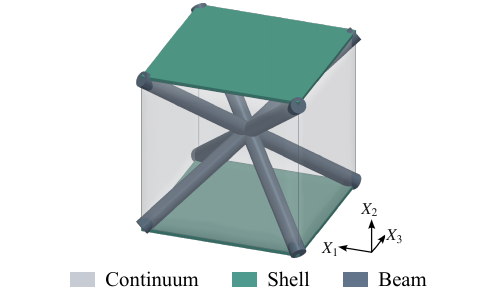}
    \caption{Complete periodic representation of the elastoplastic sandwich cell and its constituent arrangement.}
    \label{fig:sandwich_cell}%
\end{figure}

The final example retains infinitesimal kinematics while introducing history-dependent constituent behavior. The cubic cell has edge length $\ell=1$ and combines a soft elastic continuum matrix, two shell skins normal to $\boldsymbol{e}_{2}$, and a body-centered-cubic (BCC) beam core whose eight members connect the cell center to its vertices. The shell meshes conform to the matrix boundary, and the beam endpoints share the corner nodes of the continuum and shell meshes. The interior beam nodes are independently discretized and coupled to the matrix through an embedded constraint. The complete periodic representation shown in Figure~\ref{fig:sandwich_cell} contains both shell skins, whereas the ownership-consistent model removes the skin on the secondary face before solving the boundary-value problem.

All quantities are nondimensional. The matrix mesh uses the characteristic element size $\ell/8$, which is inherited by the triangular shell meshes, and each beam member is discretized by 16 elements. The matrix, shell skins, and beam core use linear tetrahedral continuum elements (C3D4), three-node shell elements (S3), and two-node Timoshenko beam elements (B31), respectively. The matrix remains linearly elastic, whereas the structural constituents follow ideal $J_{2}$ plasticity with different yield stresses $\sigma_{\mathrm{y}}$ and without microscopic hardening. Table~\ref{tab:example_3_parameters} summarizes the material and section parameters, where $t$ denotes the shell thickness and $r$ the beam radius.

\begin{table}[htbp]
    \centering
    \caption{Nondimensional constituent parameters and discretization of the elastoplastic sandwich cell.}
    \label{tab:example_3_parameters}%
    \small
    \setlength{\tabcolsep}{4.5pt}
    \renewcommand{\arraystretch}{1.1}
    \begin{tabular}{lccccc}
        \hline
        Constituent & $E$ & $\nu$ & $\sigma_{\mathrm{y}}$ & Section & Element \\
        \hline
        Matrix & $1$   & $0.3$ & --    & --       & C3D4 \\
        Shell  & $100$ & $0.3$ & $0.4$ & $t=0.02$ & S3 \\
        Beam   & $100$ & $0.3$ & $0.6$ & $r=0.05$ & B31 \\
        \hline
    \end{tabular}
\end{table}

\begin{figure}[htbp]
    \centering
    \includegraphics[width=0.95\columnwidth]{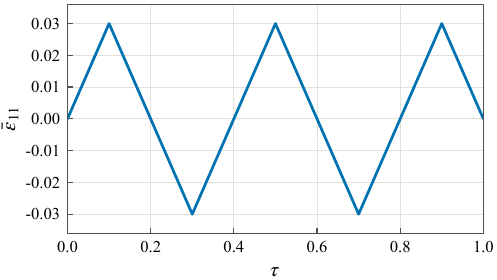}
    \caption{Prescribed macroscopic strain history for the cyclic analysis.}
    \label{fig:sandwich_load_history}%
\end{figure}

The path-dependent response is first investigated under the prescribed macroscopic strain $\bar{\varepsilon}_{11}$, while all other macroscopic strain components vanish. Following the initial loading, the piecewise-linear history in Figure~\ref{fig:sandwich_load_history} comprises two fully reversed cycles between $\bar{\varepsilon}_{11}=\pm0.03$ and a final unloading to zero strain. The magnitude of the strain rate is constant, and the quasi-static analysis uses fixed pseudo-time increments of $\Delta \tau = 0.002$ for $\tau\in[0,1]$.

\begin{figure}[htbp]
    \centering
    \includegraphics[width=0.95\columnwidth]{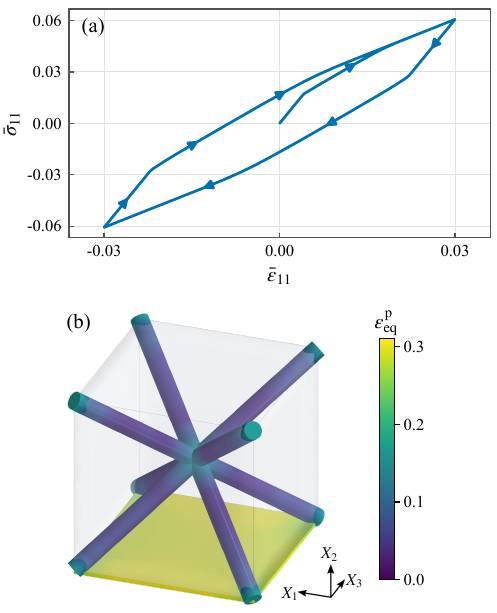}
    \caption{Cyclic response of the elastoplastic sandwich cell: (a) effective axial stress and (b) accumulated equivalent plastic strain after the final unloading.}
    \label{fig:sandwich_cyclic_response}%
\end{figure}

Figure~\ref{fig:sandwich_cyclic_response}(a) shows two axial hysteresis loops with peak stresses $\bar{\sigma}_{11} = \pm0.0607$. The second loop nearly retraces the first, indicating that the macroscopic response has stabilized at the prescribed strain amplitude. Final unloading to zero macroscopic strain leaves the residual effective stress $\bar{\sigma}_{11}=-0.0167$, reflecting the heterogeneous residual-stress state within the cell. After this unloading, the equivalent-plastic-strain field in Figure~\ref{fig:sandwich_cyclic_response}(b) reaches maxima of $0.311$ in the shell skins and $0.182$ in the beam core. Both structural phases undergo plastic flow and contribute to the effective hysteresis, with the larger plastic strains developing in the shells. Over the cyclic history, the maximum Hill--Mandel residual is $2.3 \times 10^{-5}$. The maximum relative asymmetry of the effective stress is $9.3 \times 10^{-8}$, confirming macroscopic stress symmetry also for the path-dependent response.

\begin{figure*}[htbp]
    \centering
    \includegraphics[width=0.95\textwidth]{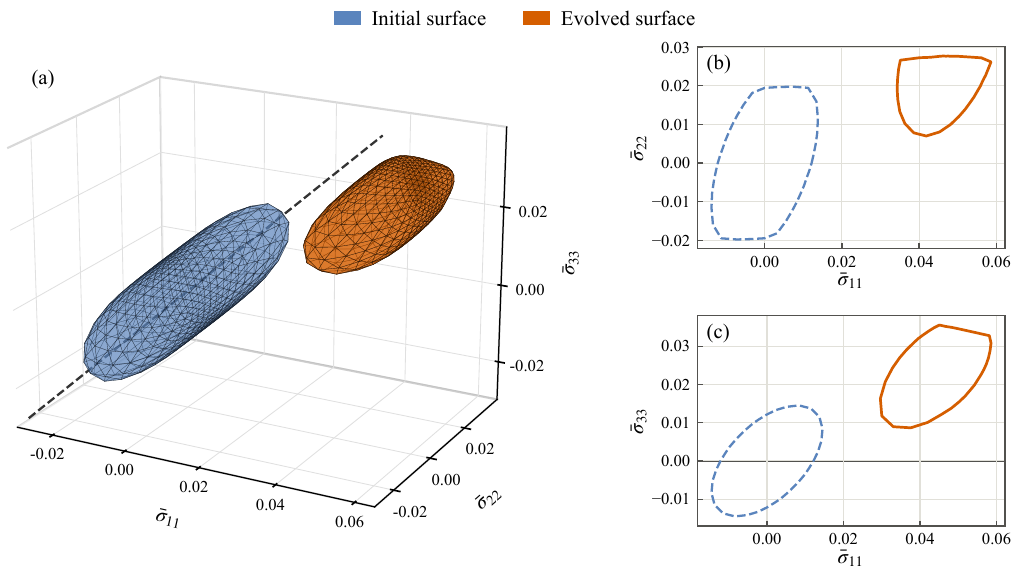}
    \caption{Initial and evolved operational yield surfaces in the normal-stress subspace: (a) three-dimensional surfaces with the hydrostatic axis $\bar{\sigma}_{11}=\bar{\sigma}_{22}=\bar{\sigma}_{33}$ indicated by the dashed line. Independently sampled sections through the respective probing states are shown in (b) the $\bar{\sigma}_{11}$--$\bar{\sigma}_{22}$ plane and (c) the $\bar{\sigma}_{11}$--$\bar{\sigma}_{33}$ plane.}
    \label{fig:sandwich_yield_surfaces}%
\end{figure*}

To characterize the history-induced evolution of the elastic domain, the operational yield surface is sampled in the three-dimensional subspace of normal effective stresses $(\bar{\sigma}_{11},\bar{\sigma}_{22},\bar{\sigma}_{33})$. The initial and evolved surfaces are each computed in 642 directions distributed over the unit sphere in the three-dimensional normal-stress space. The initial surface is obtained from the stress-free cell. The evolved surface is evaluated after monotonic loading to $\bar{\varepsilon}_{11} = 0.03$ and elastic unloading to a common state inside the resulting elastic domain. Every evolved-state probe reproduces this history before loading in the selected direction. Two-dimensional sections through the respective probing states and parallel to the $\bar{\sigma}_{11}$--$\bar{\sigma}_{22}$ and $\bar{\sigma}_{11}$--$\bar{\sigma}_{33}$ planes are additionally sampled directly using 36 directions each. Because the probes are strain controlled, the prescribed stress-increment directions are mapped to proportional strain-increment directions using the numerically determined effective elastic compliance. The probe paths use fixed increments satisfying $\lVert \Delta \bar{\boldsymbol{\varepsilon}}^{\mathrm{V}} \rVert_{\infty} \leq 6 \times 10^{-5}$. Motivated by dissipation-based operational definitions of macroscopic yielding in unit-cell simulations \citep{chouksey_2019, chouksey_2021}, the present yield indicator is
\begin{align}
    \chi_{\mathrm{p}} \coloneqq \frac{\Delta W_{\mathrm{p}}}{\left\lvert \Delta W_{\mathrm{int}} \right\rvert} \,,
\end{align}
where $\Delta W_{\mathrm{p}}$ and $\Delta W_{\mathrm{int}}$ are the changes in plastic dissipation and internal energy from the beginning of the respective probe. Operational yielding is identified by linearly interpolating the first crossing of $\chi_{\mathrm{p}} = 10^{-4}$. An increase in the maximum equivalent plastic strain over all shell and beam integration points is additionally required to confirm the onset of plastic flow.

Figure~\ref{fig:sandwich_yield_surfaces}(a) shows that the independently sampled initial surface is centrally symmetric. It is centered at the origin and elongated approximately along the hydrostatic axis, but remains bounded in that direction. The effective onset of yielding is therefore pressure sensitive even though the structural constituents obey pressure-insensitive $J_{2}$ plasticity. A macroscopic hydrostatic stress state generally produces heterogeneous, non-hydrostatic local stresses in the mixed-dimensional microstructure. After tensile preloading, the surface is displaced from the origin toward the tensile prestressed state and changes both size and shape. The three-dimensional comparison in Figure~\ref{fig:sandwich_yield_surfaces}(a) and the independently sampled sections in Figures~\ref{fig:sandwich_yield_surfaces}(b) and (c) show that this evolution is direction dependent. The volume enclosed by the triangulated three-dimensional surface decreases to $46.7\,\%$ of its initial value. The polygonal areas of the $\bar{\sigma}_{11}$--$\bar{\sigma}_{22}$ and $\bar{\sigma}_{11}$--$\bar{\sigma}_{33}$ sections decrease to $44.2\,\%$ and $92.9\,\%$, respectively. The evolution can therefore be represented neither by uniform expansion nor by rigid translation. Instead, it combines a Bauschinger-type shift with nonuniform contraction and distortion. Plastic flow resumes during the first reversal in Figure~\ref{fig:sandwich_cyclic_response}(a) while the axial effective stress remains tensile, consistent with the shifted evolved surface. Because the structural constituents are ideally plastic, the effective evolution results from heterogeneous plastic strain and residual stress within the RVE rather than from microscopic hardening.

% ========================================================================================
\section{Conclusion and outlook}\label{sec:conclusion}%
This work establishes a common nonlinear first-order computational homogenization framework for periodic RVEs assembled from compatible continuum, truss, beam, and shell constituents. Rather than defining the effective medium, an auxiliary finite-strain Cosserat continuum provides their common power-equivalent representation of virtual power and generalized boundary kinematics. The uniform rotation factor of this auxiliary representation is identified with the macroscopic polar rotation, and its associated macroscopic wryness is zero, so the macroscale body remains a conventional Cauchy continuum while microscopic rotations and moments remain in the RVE equilibrium. The resulting periodic translational and rotational conditions and nodal-force recovery of the effective first Piola--Kirchhoff stress are independent of the constituent formulation. Microscopic moments affect the macroscopic response through the boundary forces rather than an independent couple-stress measure.

For lower-dimensional members supported on periodically identified cell boundaries, fractional ownership is assigned through the internal virtual work and shown to be variationally equivalent to retaining one representative of every set of periodic element copies while preserving the complete nodal topology. The equivalence extends to the residual and its consistent linearization, allowing geometrically nonlinear, materially nonlinear, and path-dependent formulations without rescaling section properties, constitutive parameters, or internal variables. Together, the common scale transition, ownership-consistent reduction, and nodal-force stress recovery provide a single procedure for all supported constituent types, implemented in the open-source library \texttt{abqhom}.

The three numerical examples examine complementary aspects of the framework. The mixed continuum--shell--beam cell verifies the analytical small-strain stiffness, the equivalence between weighted and reduced representations, and the discrete power balance. The hyperelastic Octet lattice demonstrates finite rotations and an imperfection-triggered buckling response, whereas the elastoplastic sandwich cell exhibits macroscopic hysteresis and history-dependent evolution of its effective yield surface despite ideal plasticity of its structural constituents. These results show that the same scale transition remains applicable across distinct element kinematics and constitutive regimes. The present setting is restricted to first-order homogenization of periodic cuboidal cells with matching discretizations on paired cell boundaries and consequently does not resolve macroscopic size effects. Extensions to nonmatching periodic boundary discretizations, non-orthogonal periodic cells, refined joint models, and fully nested FE\textsuperscript{2} simulations based on a consistently condensed macroscopic tangent provide natural directions for future work.

% ========================================================================================
\section*{CRediT authorship contribution statement}
\textbf{Jonathan Stollberg:} Conceptualization, Methodology, Software, Formal analysis, Investigation, Writing -- original draft, Visualization. \textbf{Dominik Schillinger:} Writing -- review \& editing, Supervision, Project administration, Funding acquisition.

\section*{Data availability}
The source code of \texttt{abqhom} and the scripts required to reproduce the numerical examples are publicly available at \href{https://github.com/jstollberg/abqhom}{github.com/jstollberg/abqhom}. A versioned release of the repository will be archived on Zenodo prior to publication, and the corresponding DOI will be added to the final manuscript.

\section*{Declaration of competing interest}
The authors declare that they have no known competing financial interests or personal relationships that could have appeared to influence the work reported in this paper.

\section*{Funding}
Open access funding enabled and organized by Projekt DEAL.

\section*{Declaration of generative AI and AI-assisted technologies in the manuscript preparation process}
During the preparation of this work, the authors used GPT-5.6 Sol and Claude Opus 4.8 in order to improve the grammar, language, and readability of the manuscript. The tools also assisted with the review, refactoring, testing, and documentation of the \texttt{abqhom} source code and example scripts. After using these tools, the authors reviewed and edited the content as needed and take full responsibility for the content of the published article.

% ========================================================================================
\appendix
\section{Symmetry of the effective Cauchy stress}\label{app:symmetry}%
Following the argument of \citet{miehe_2004}, this appendix proves that the effective Cauchy stress~\eqref{eq:cauchy_from_P} obtained under periodic boundary conditions is symmetric, $\bar{\boldsymbol{\sigma}} = \bar{\boldsymbol{\sigma}}^{\mathrm{T}}$, equivalently that the skew part of the macroscopic Kirchhoff stress $\bar{\boldsymbol{P}} \bar{\boldsymbol{F}}^{\mathrm{T}}$ vanishes.

Multiplying the force virial~\eqref{eq:mean_stress} by $\bar{\boldsymbol{F}}^{\mathrm{T}}$, using the dyad identity $(\boldsymbol{t} \otimes \boldsymbol{X}) \bar{\boldsymbol{F}}^{\mathrm{T}} = \boldsymbol{t} \otimes (\bar{\boldsymbol{F}} \boldsymbol{X})$, and inserting the placement localization from \eqref{eq:first_order_localization} gives
\begin{align}
    \bar{\boldsymbol{P}} \bar{\boldsymbol{F}}^{\mathrm{T}} = \frac{1}{V} \int_{\partial \Omega_{0}} \boldsymbol{t} \otimes \boldsymbol{x} \,\mathrm{d} A - \frac{1}{V} \int_{\partial \Omega_{0}} \boldsymbol{t} \otimes \tilde{\boldsymbol{u}} \,\mathrm{d} A \,.
\end{align}
The fluctuation integral vanishes because $\tilde{\boldsymbol{u}}$ is periodic while $\boldsymbol{t}$ is anti-periodic, so the contributions of paired faces cancel, exactly as in the fluctuation-work cancellation~\eqref{eq:zero_boundary_fluctuation_work}. This leaves the boundary virial of the macroscopic Kirchhoff stress,
\begin{align}
    \bar{\boldsymbol{P}} \bar{\boldsymbol{F}}^{\mathrm{T}} = \frac{1}{V} \int_{\partial \Omega_{0}} \boldsymbol{t} \otimes \boldsymbol{x} \,\mathrm{d} A \,.
    \label{eq:macro_kirchhoff_boundary_virial}%
\end{align}
Taking the axial vector of its skew part and using the identity $2 \operatorname{axl} \bigl( \operatorname{skew}( \boldsymbol{a} \otimes \boldsymbol{b} ) \bigr) = \boldsymbol{b} \times \boldsymbol{a}$, which holds for any vectors $\boldsymbol{a}, \boldsymbol{b}$ under the convention~\eqref{eq:hat_operator_definition}, turns \eqref{eq:macro_kirchhoff_boundary_virial} into the mean moment of the boundary force tractions,
\begin{align}
    2 \operatorname{axl} \bigl( \operatorname{skew}( \bar{\boldsymbol{P}} \bar{\boldsymbol{F}}^{\mathrm{T}} ) \bigr) = \frac{1}{V} \int_{\partial \Omega_{0}} \boldsymbol{x} \times \boldsymbol{t} \,\mathrm{d} A \,.
    \label{eq:macro_stress_skew_moment}%
\end{align}

It remains to show that this mean moment vanishes. Applying the divergence theorem to the moment of the force tractions $\boldsymbol{t} = \boldsymbol{P} \boldsymbol{N}$ and using the force balance~\eqref{eq:force_balance} together with $\operatorname{Grad} \boldsymbol{x} = \boldsymbol{F}$ and the definition~\eqref{eq:internal_moment_definition} of the internal moment vector $\boldsymbol{r}$ gives
\begin{align}
    \int_{\partial \Omega_{0}} \boldsymbol{x} \times \boldsymbol{t} \,\mathrm{d} A = \int_{\Omega_{0}} \boldsymbol{r} \,\mathrm{d} V \,.
    \label{eq:force_moment_to_internal}%
\end{align}
Likewise, the divergence theorem applied to the couple tractions $\boldsymbol{c} = \boldsymbol{\mu} \boldsymbol{N}$ with the couple balance~\eqref{eq:couple_balance} gives
\begin{align}
    \int_{\partial \Omega_{0}} \boldsymbol{c} \,\mathrm{d} A = \int_{\Omega_{0}} \operatorname{Div} \boldsymbol{\mu} \,\mathrm{d} V = - \int_{\Omega_{0}} \boldsymbol{r} \,\mathrm{d} V \,.
    \label{eq:couple_moment_to_internal}%
\end{align}
Adding \eqref{eq:force_moment_to_internal} and \eqref{eq:couple_moment_to_internal} eliminates the internal moment and recovers the global angular-momentum balance of the unit cell,
\begin{align}
    \int_{\partial \Omega_{0}} \bigl( \boldsymbol{x} \times \boldsymbol{t} + \boldsymbol{c} \bigr) \,\mathrm{d} A = \boldsymbol{0} \,.
\end{align}
Under the periodic boundary conditions the couple tractions are anti-periodic, so their resultant cancels pairwise. Equation~\eqref{eq:force_moment_to_internal} then shows that the mean internal moment vanishes as well, and the moment of the force tractions reduces to
\begin{align}
    \int_{\partial \Omega_{0}} \boldsymbol{x} \times \boldsymbol{t} \,\mathrm{d} A = \boldsymbol{0} \,.
\end{align}
Substituting this into \eqref{eq:macro_stress_skew_moment} yields
\begin{align}
    \operatorname{skew}( \bar{\boldsymbol{P}} \bar{\boldsymbol{F}}^{\mathrm{T}} ) = \boldsymbol{0} \,,
\end{align}
hence $\bar{\boldsymbol{\sigma}} = \bar{\boldsymbol{\sigma}}^{\mathrm{T}}$, as required for the first-order Cauchy macro-continuum.

\bibliographystyle{elsarticle-num-names}
\bibliography{bibliography}

\end{document}